\documentclass[review]{elsarticle}

\usepackage{hyperref}
\usepackage{amssymb,amsmath,amsthm}
\usepackage{subcaption,caption}
\usepackage{appendix}
\usepackage{color}
\usepackage{booktabs}
\usepackage{threeparttable}
\usepackage{fullpage}

\def\figwidth{0.8\textwidth}
\begin{document}

\begin{frontmatter}

\title{Exact mean-field models for spiking neural networks with adaptation\tnoteref{mytitlenote}}
\tnotetext[mytitlenote]{This work was supported by the Natural Sciences and Engineering Research Council of Canada.}

\author{L. Chen}
\ead{L477chen@uwaterloo.ca}

\author{S. A. Campbell\corref{mycorrespondingauthor}}
\cortext[mycorrespondingauthor]{Corresponding author}
\ead{sacampbell@uwaterloo.ca}

\address{Department of Applied Mathematics and Centre for Theoretical Neuroscience, University of Waterloo, Waterloo, ON, N2L 3G1, Canada}

\begin{abstract}
Networks of spiking neurons with adaption have been shown to be able to reproduce a wide range of neural activities, including the emergent population bursting and spike synchrony that underpin brain disorders and normal function. 
Exact mean-field models derived from spiking neural networks are extremely valuable, as such models can be used to determine how individual neuron and network parameters interact to produce macroscopic network behaviour. 
In the paper, we derive and analyze a set of exact mean-field equations for the neural network with spike frequency adaptation. Specifically, our model is a network of Izhikevich neurons, where each neuron is modeled by a two dimensional system consisting of a quadratic integrate and fire equation plus an equation which implements spike frequency adaptation. 
Previous work deriving a mean-field model for this type of network, relied on the assumption of sufficiently slow dynamics of the adaptation variable.
However, this approximation did not succeeded in establishing an exact correspondence between the macroscopic description and the realistic neural network, especially when the adaptation time constant was not large.
The challenge lies in how to achieve a closed set of mean-field equations with the inclusion of the mean-field expression of the adaptation variable.
We address this challenge by using a Lorentzian ansatz combined with the moment closure approach to arrive at the mean-field system in the thermodynamic limit.
The resulting macroscopic description is capable of qualitatively and quantitatively describing the collective dynamics of the neural network, including transition between tonic firing and bursting.
We extend the approach to a network of two populations of neurons and discuss the accuracy and efficacy of our mean-field approximations by examining all assumptions that are imposed during the derivation.
Numerical bifurcation analysis of our mean-field models reveal bifurcations not previously observed in the models, including a novel mechanism for emergence of bursting in the network. 
We anticipate our results will provide a tractable and reliable tool to investigate the underlying mechanism of brain function and dysfunction from the perspective of theoretical neuroscience.

\end{abstract}


\end{frontmatter}

%\linenumbers

\section{INTRODUCTION}
A central topic of theoretical neuroscience research is to obtain computationally and/or analytically tractable models for understanding the neural dynamics that underpin brain disorders, such as  epilepsy or Parkinson’s disease, or for normal functioning, such as memory or decision making.
The dynamics associated with such functions or disorders  result from the coordinated activity of large populations of interconnected neurons. 
Neural mass models, rooted in the mean field theory, aim to describe the collective activity of a neural network in terms of mean-field variables such as the population firing rate and mean membrane potential.
The development of mean-field models have been a long history spanning more than a half-century \cite{Deco2008}. In the current literature, there are two common types of model.  One is based on  heuristic descriptions that are designed to resemble macroscopic features of neural dynamics based on physiological observations and disregard individual behaviours of the neural network. The Wilson-Cowan model \cite{WilsonCowan2021} is arguably the most influential one. 
As an alternative, exact macroscopic descriptions have been developed through mean-field reduction of   networks of model neurons by using concepts from statistical physics. These models bridge the microscopic properties of individual neurons and macroscopic collective dynamics of the neural network.  Thus they can account for phenomena that the heuristic models cannot, such as synchronization mechanisms that arise due to the interaction between individual neuron behaviour and network properties.
The development of exact mean-field models has followed two threads, both based on the population density approach from statistical physics. This approach yields a conservation law for the population density function, which exactly represents the dynamics of the network in the limit that the number of elements becomes arbitrarily large. 
In one thread, this approach was applied to networks of spiking neurons \cite{abbott1993asynchronous,treves1993mean,nykamp2000population,omurtag2000simulation,apfaltrer2006population}. In the other thread, this approach was applied to networks of coupled phase models \cite{strogatz1991stability,watanabe1993integrability,watanabe1994constants,Ott2008,pikovsky2008,Ott2009,Ott2011}.
Recently, these two threads have started to converge, when it was shown that the approach of Ott and Antonsen \cite{Ott2008,Ott2009,Ott2011} could be applied to networks of coupled Quadratic Integrate-and-Fire (QIF) neurons using the link between the QIF model and the theta model \cite{Montbrio2015}.

Spiking neural models involve variables closely related to biological measurements and realizations. The QIF model is a popular model for large network simulations as it only has one equation for each neuron. Further, it can be considered a canonical model as any Class I excitable system close enough to the onset of oscillations can be transformed into this form  \cite{ermentrout1986parabolic,Ermentrout1996,Latham2000}. Following the work of \cite{Montbrio2015}, recent work has focussed on developing mean-field models for 
%to study the collective neurodynamics of 
QIF networks with 
added biophysical mechanisms or structural details to explain interesting neural activities. 
These additions include the incorporation of propagation delays as action potentials travel along axons or dendrites \cite{Ratas2018}, gap junctions between neurons \cite{Montbrio2020} and the short-term synaptic plasticity \cite{Gast2021PhyRevE}. More information on mean-field models of such extended neural networks, can be found in recent reviews \cite{Ashwin2016Rew,Bick2020}.

While the QIF model is useful, there are many behaviours of spiking neurons it cannot reproduce. To address this, several authors have developed two-dimensional integrate-and-fire model neurons. Examples include  the Izhikevich neuron \cite{Izhikevich2003} and the adaptive exponential (AdEx) neuron \cite{Brette2005_AdEx}.  These models
display spike frequency adaptation (SFA) through a recovery variable and 
are capable of generating a variety of spiking dynamics reported in real neurons \cite{Izhikevich2004, Touboul2008, Wulfram2014}. 
The SFA mechanism 
can improve neural coding and computation at a lower
metabolic cost \cite{Gutierrez2019,Fitz2020,Salaj2021} and has also been demonstrated to be important in the emergence of network bursting and synchronization \cite{vreeswijk2001patterns,nesse2008fluctuation,KilpatrickErmentrout2011,ferguson2015,Gast2020NeuralComp}. 
Thus, derivation of mean-field descriptions for  networks of neurons with SFA would be extremely valuable.

Since the Izhikevich neuron is the most closely linked to the QIF neuron, it is the ideal candidate to explore emergent neurodynamics of neural networks through mean-field descriptions.
Further, the Izhikevich network has been widely employed to study brain function, e.g., \cite{izhikevich2008large,Dur-e-Ahmad2012} and dysfunction, e.g., \cite{Rich2020}. 
The Izhikevich neuron model consists of a fast subsystem based on the QIF model and a slow subsystem modelling the adaptation mechanism. Thus the Izhikevich network can be considered as a QIF network extended by SFA. In this article, we show it is feasible to extend the Lorentzian ansatz \cite{Montbrio2015}, or the equivalent OA ansatz \cite{Ott2008} for the phase model, to the derivation of the exact mean-field models for a network of Izhikevich neurons. 
The difficulty lies in how to achieve a closed set of mean-field equations with the inclusion of the mean-field expression of the adaptation variable. To do this we turn to the  population density approach for spiking neurons, %\cite{abbott1993asynchronous,treves1993mean,nykamp2000population,omurtag2000simulation,apfaltrer2006population}, 
which was extended to two dimensional integrate-and-fire models by Nicola and Campbell
\cite{Nicola2013bif,Nicola2013hetero}.
The quasi-steady approximation for the continuity equation in \cite{Nicola2013hetero} will be replaced with the Lorentzian ansatz \cite{Montbrio2015} to drop the assumption of separation of time scales. The moment closure approach \cite{Ly2007} will be deployed to release the dependence of the adaptation variable on the membrane potential to help close the mean-field system.

We apply our method to two examples. The first consists of a single population of identical neurons with all-to-all coupling and heterogenous input currents. We confirm that the bifurcation structure is similar to that observed in \cite{Nicola2013hetero}. In our model, the frequency mismatch between the network and the mean-field model which occurs in \cite{Nicola2013hetero} no longer exists. Further, we give evidence for a co-dimension two bifurcation that organizes the system behaviour.
The second model is a network consisting of two populations of neurons. Both populations are all-to-all coupled internally and externally. One population has strong SFA while the other has weak SFA. We explore the bifurcations that occur as the relative proportions of the two neurons is varied and show that multiple co-dimension two bifurcations occur. One interesting consequence is that the onset of bursting in the network can change from a subcritical Hopf bifurcation of the mean firing rate to a homoclinic bifurcation or saddle-node on an invariant circle bifurcation. 

The paper is structured as follows. In section~\ref{sec:model}, we introduce the specific neuron model and network which will be used throughout. 
In section~\ref{sec:derivation}, we present a detailed derivation of the mean-field equations for a single population. The assumptions are added on one after the other to show their role in forming the final mean-field model. 
In section~\ref{sec:numerics}, we consider the single population example. We perform a bifurcation analysis of the mean-field system and show, by comparison with numerical simulation of a finite-size neural network, that the mean-field model has excellent agreement with the network behaviour. 
In section~\ref{sec:2pop}, we extend the model to a network consisting of two populations of neurons and then apply this to the example
described above. %Both populations are all-to-all coupled internally and externally.  
Numerical simulations and numerical bifurcation diagrams demonstrate the validity of the developed mean-field description. 
Finally, in section~\ref{sec:discussion} we conclude the paper by discussing the influence of assumptions required in our approach on the accuracy and efficacy of the mean-field models, and what can be done to extend this approach when these assumptions  cannot be met.

%%%%%%%%%%%%%%%%%%%%%%%%%%%%%%%%%%%%%

\section{THE NETWORK SYSTEM}\label{sec:model}
The Izhikevich model is a result of reduction of the biophysically accurate Hodgkin-Huxley type neuron model through bifurcation analysis \cite{Izhikevich2007}. It still retains central properties of neural activity with adaptive quadratic mechanism. The network model for a population of Izhikevich neurons is described by the following discontinuous ordinary differential equations (ODEs),
\begin{subequations}\label{eq:network-Izh}
\begin{align}
    v'_k 
    &=
    v_k(v_k - \alpha)
    -
    w_k
    +
    \eta_k + I_{\mathrm{ext}}
    +
    I_{\mathrm{syn},k} 
    \\
    w'_k
    &=
    a(b v_k - w_k)
    \\
    \mathrm{if} \;
    &
    v_k \geq v_{\mathrm{peak}}, 
    \mathrm{then} \;
    v_k \leftarrow v_{\mathrm{reset}}\;
    \mathrm{and}\;
    w_k \leftarrow w_k + w_{\mathrm{jump}}
\end{align}
\end{subequations}
for $k=1, 2, \dots, N$. Here, $'=d/dt$ denotes the time derivative, $v_k(t)$ is the membrane potential of $k$th neuron and $w_k$ is the recovery current, which serves as an adaptation variable. By construction, $v_k\in [v_\mathrm{reset}, v_\mathrm{peak}]$, while $w$ has no constraints, $w_k\in (-\infty,\infty)$. The parameter $\eta_k$ is the intrinsic current while $I_\mathrm{ext}$ is the external common current. We will assume that $\eta_k$ are heterogeneous and drawn from a distribution ${\mathcal L}(\eta)$ defined on $(-\infty,\infty)$. The term $I_{\mathrm{syn}}$ represents the total synaptic current due to the other neurons in the network.
When the voltage reaches a cut off value $v_{\mathrm{peak}}$, considered to be the peak of a spike, it is reset to the value $v_{\mathrm{reset}}$. At the same time, the adaptation variable jumps by an amount $w_{\mathrm{jump}}$, which affects
%a total current affecting 
the after spike behavior. 
When $v_{\mathrm{peak}} = -v_{\mathrm{reset}}$ approaches $\infty$, the Izhikevich model (\ref{eq:network-Izh}) can be transformed into the theta model with adaptation, where the neuron fires a spike whenever $\theta$ crosses $\pi$  \cite{Izhikevich2000}.

Neurons in the network are connected by synapses modeled by
\begin{equation}
    I_{\mathrm{syn},k} = g_{\mathrm{syn}}s(e_r - v_k),
\end{equation}
where $e_r$ the reversal potential and $g$ is the maximum synaptic conductance, both assumed to be the same for all neurons.   The synaptic gating variable, $s$, lies between 0 to 1, and represents the proportion of ion channels open in the postsynaptic neuron as the result of the firing in presynaptic neurons. For a network with all-to-all connectivity, $s$ is homogeneous across the network as every post synaptic neuron receives the same summed input from all the presynaptic neurons.
The mechanism of synaptic transimission can be formally described by a linear system of ODEs with a sum of delta pulses corresponding to the times a neuron fires a spike. For example, the single exponential synapse is modeled by
\begin{equation}\label{eq:single-exp-synapse}
s'=
-\frac{s}{\tau_s}
+
\frac{s_{\mathrm{jump}}}{N}
\sum_{k=1}^{N}
\sum_{t_k^j<t}
\delta(t-t_k^j),     
\end{equation}
where $\delta(t)$ is the Dirac delta function, and $t_k^j$ represents the time of the $j$th spike of the $k$th neuron. The double exponential synapse and the alpha synapse are also frequently used in the literature. Their dynamics can be described by two-coupled first-order ODEs \cite{Ermentrout2010book}. For simplicity, we consider the single exponential synaptic dynamics (\ref{eq:single-exp-synapse}) in this paper. It is easy to extend our work to other synaptic models, see  \cite{Nicola2013bif}. We also assume all-to-all connectivity and that the synaptic parameters $s_{\mathrm{jump}}$ and $\tau_s$ are the same for every synapse. 

Here, the network system of (\ref{eq:network-Izh})-(\ref{eq:single-exp-synapse}) is dimensionless, which is appropriate for mathematical and numerical exploration of the neurodynamics. However, neuroscientists are normally accustomed to the dimensional form with parameters that have physiological interpretation, e.g., \cite{Dur-e-Ahmad2012} and \cite{Rich2020}.  
So we present in \ref{app:nondim} the details of transformation between the dimensional and dimensionless models.

%%%%%%%%%%%%%%%%%%%%%%%%%%%%%%%%%%%%%

\section{MEAN-FIELD REDUCTION}\label{sec:derivation}
The network model described in the previous section is too complicated to perform tractable analysis especially when the number of neurons is large. In this section, we will develop a low-dimensional mean-field model to approximate the behaviour of the full network described by (\ref{eq:network-Izh})-(\ref{eq:single-exp-synapse})  within the thermodynamic limit, i.e., when $N \rightarrow \infty$. 

The mean-field approximation is essentially a technique that borrows concepts and methods from statistical physics, e.g., the population density approach \cite{Ly2007}, the continuity equation (or the Fokker-Planck equation when the system is subject to noise) \cite{Deco2008,Bick2020}.
We will show how to describe some vital macroscopic variables such as the population firing rate 
%by deploying statistical terminology 
and how to derive the reduced macroscopic dynamics, cast as ODEs, through step-by-step assumptions. Our approach comes from combining the ideas of \cite{Nicola2013bif,Nicola2013hetero, Ly2007} and \cite{Montbrio2015}.

\subsection{General mean-field description}
We define the population density function $\rho(t,v,w,\eta)$ as the density of neurons at a point $(v,w)$ in phase space and parameter $\eta$ at time $t$. In the limit as $N \rightarrow \infty$, the principle of conservation mass leads to the following evolution equation for the density function, that is, the continuity equation,
\begin{equation}\label{eq:CE-general}
    \frac{\partial}{\partial t}
     \rho(t,v,w,\eta)
    +
    \bigtriangledown \cdot \mathcal{J}(t,v,w,s,\eta)
    =0,
\end{equation}
where the probability flux is defined as
\begin{align}\label{eq:flux-general}
  \mathcal{J}(t,v,w,s,\eta) 
  &= 
  \begin{pmatrix}
  \mathcal{J}^v(t,v,w,s,\eta) \\
  \mathcal{J}^w(t,v,w)
  \end{pmatrix} \nonumber\\
  &= 
  \begin{pmatrix}
  G^v(v,w,s,\eta)\\
  G^w(v,w)
  \end{pmatrix}
  \rho(t,v,w,\eta) \nonumber\\
  &= 
  \begin{pmatrix}
  v(v-\alpha) - w + \eta + I_{\mathrm{ext}} +g_{\mathrm{syn}}s(e_r - v) \\
  a(bv-w)
  \end{pmatrix}
  \rho(t,v,w,\eta).
\end{align}
Note that $\mathcal{J}^v(t,v,w,s,\eta)$ is $s$ dependent. The flux is intuitively the mass flow rate along a specific direction in phase space.
A boundary condition for the flux, consistent with the resetting rule in (\ref{eq:network-Izh}), is imposed,
\begin{equation}\label{eq:boundary}
    \mathcal{J}^v(t,v_{\mathrm{peak}},w,s,\eta)
    =
    \mathcal{J}^v(t,v_{\mathrm{reset}},w+w_{\mathrm{jump}},s,\eta).
\end{equation}
We assume the flux to be vanishing on the boundary $\partial w$ \cite{Ly2007}, i.e., in the limit $w\rightarrow \pm \infty$.

Next, we describe several macroscopic observables in terms of mean-field description, which are extremely useful in understanding neural activities underlying brain function.
The population firing rate is the flux through the threshold $v_{\mathrm{peak}}$ over the entire range of $w$ in phase space and $\eta$ in parameter space, defining
\begin{align}\label{eq:r-general}
r(t) 
&=
\lim_{v \rightarrow v_{\mathrm{peak}}}
\int_\eta \int_w 
\mathcal{J}^v(t,v,w,s,\eta) 
\mathrm{d}w
\mathrm{d}\eta \nonumber \\
& \equiv
\int_\eta \int_w 
\mathcal{J}^v(t,v_{\mathrm{peak}},w,s,\eta) 
\mathrm{d}w
\mathrm{d}\eta
\end{align}
The mean membrane potential is defined as 
\begin{equation}\label{eq:v-general}
\langle v(t) \rangle =
\int_\eta \int_w \int_v
v \rho(t,v,w,\eta)
\mathrm{d}v
\mathrm{d}w
\mathrm{d}\eta
\end{equation}
where $\langle \cdot \rangle$ represent the average over the population.
Additionally, we define the mean adaptation current over the population as
\begin{equation}
\langle w(t) \rangle =
\int_\eta \int_v \int_w
w \rho(t,v,w,\eta)
\mathrm{d}w
\mathrm{d}v
\mathrm{d}\eta
\end{equation}
Then, we approximate its derivative with respect to $t$, yielding
\begin{align}
\langle w \rangle' 
&= 
\int_\eta \int_v \int_w
w 
\frac{\partial}{\partial t}\rho(t,v,w,\eta)
\mathrm{d}w
\mathrm{d}v
\mathrm{d}\eta \nonumber\\
& \approx
\langle G^w(v,w) \rangle
+
\int_\eta \int_w
w_{\mathrm{jump}}
\mathcal{J}^v(t,v_{\mathrm{peak}},w,s,\eta)
\mathrm{d}w
\mathrm{d}\eta
\end{align}
To obtain this expression, we assume $\langle w|\eta \rangle \gg w_{\mathrm{jump}}$ and  the flux to be vanishing on the boundary $\partial w$ \cite{Nicola2013bif}. See \ref{app:dw}  for more details.
Further, considering the linearity of $G^w(\cdot)$  function with respect to $v$ and $w$, see eq. (\ref{eq:flux-general}), and the description of the population firing rate  in terms of flux (\ref{eq:r-general}), we finally derive the following ODE describing the evolution of the mean adaptation variable,
\begin{align}\label{eq:MF-w}
\langle w \rangle' 
&= 
G^w (\langle v \rangle, \langle w \rangle)
+ w_{\mathrm{jump}} r(t) \nonumber \\
&=
a
\left (
b \langle v \rangle - \langle w \rangle
\right )
+ w_{\mathrm{jump}} r(t).
\end{align}
By considering the relationship between the flux and the description of the population firing rate in terms of number of spikes fired by neurons \cite{Nicola2013bif}, we also can rewrite the synaptic dynamics (\ref{eq:single-exp-synapse}) in terms of the firing rate as
\begin{align}\label{eq:MF-s}
s' 
&=
-\frac{s}{\tau_s}
+
s_{\mathrm{jump}}
\int_\eta \int_w
\mathcal{J}^v(t,v_{\mathrm{peak}},w,s,\eta)
\mathrm{d}w\mathrm{d}\eta \nonumber \\
&=
-\frac{s}{\tau_s}
+
s_{\mathrm{jump}} r(t)
\end{align}
The two equations (\ref{eq:MF-w}) and (\ref{eq:MF-s}) are an integral part of the final mean-field model for the network of Izhikevich neurons. They depend on two macroscopic variables:  the mean membrane potential $\langle v(t) \rangle$ and the population firing rate $r(t)$. In the following, we will derive the dynamical system for these two variables.

%===============================
\subsection{Density function in conditional form}
In this section, we take advantage of the population density approach and the moment closure assumption to reduce the dependence between the macroscopic variables.
We begin by writing out the population density function in the conditional form
\begin{equation}\label{eq:pop-density}
    \rho (t,v,w,\eta) = \rho^w(t,w|v,\eta)\rho^v(t,v|\eta)\mathcal{L}(\eta).
\end{equation}
The population firing rate in the general expression (\ref{eq:r-general}) is then described by the conditional probability $\rho^v(t,v|\eta)$ as 
\begin{align}
r(t) 
&=
\lim_{v\rightarrow v_{\mathrm{peak}}}
\int_\eta \int_w 
\mathcal{J}^v(t,v,w,s,\eta) \mathrm{d}w \nonumber\\
&=
\lim_{v\rightarrow v_{\mathrm{peak}}}
\int_\eta \int_w
G^v(v,w,s,\eta)
\rho(t,v,w,\eta)
\mathrm{d}w
\mathrm{d}\eta \nonumber \\
&=
\lim_{v\rightarrow v_{\mathrm{peak}}}
\int_\eta \int_w
G^v(v,w,s,\eta)
\rho^w(t,w|v,\eta)\rho^v(t,v|\eta)\mathcal{L}(\eta)
\mathrm{d}w
\mathrm{d}\eta \nonumber \\
&=
\lim_{v\rightarrow v_{\mathrm{peak}}}
\int_\eta \mathcal{L}(\eta)\rho^v(t,v|\eta)
\int_w 
G^v(v,w,s,\eta)\rho^w(t,w|v,\eta)
\mathrm{d}w
\mathrm{d}\eta \nonumber \\
&=
\lim_{v\rightarrow v_{\mathrm{peak}}}
\int_\eta \mathcal{L}(\eta)\rho^v(t,v|\eta)
G^v(v,\langle w|v,\eta \rangle,s,\eta)
\mathrm{d}\eta.
\end{align}
Next, we assume 
\begin{equation}\label{eq:mc1}
\langle w|v,\eta \rangle = \langle w|\eta \rangle,    
\end{equation}
which corresponds to a first order moment closure assumption \cite{Nicola2013hetero}. Then, we have
\begin{equation}\label{eq:r-cond}
r(t) = 
\lim_{v\rightarrow v_{\mathrm{peak}}}
\int_\eta \mathcal{L}(\eta)\rho^v(t,v|\eta)
G^v(v,\langle w|\eta \rangle,s,\eta)
\mathrm{d}\eta.
\end{equation}
Similarly, the mean membrane potential is rewritten as
\begin{align}\label{eq:mean-v-cond}
\langle v(t) \rangle
&=
\int_\eta \int_w \int_v
v \rho(t,v,w,\eta)
\mathrm{d}v
\mathrm{d}w
\mathrm{d}\eta \nonumber\\
&=
\int_\eta \int_w \int_v
v \rho^w(t,w|v,\eta)\rho^v(t,v|\eta)\mathcal{L}(\eta)
\mathrm{d}v
\mathrm{d}w
\mathrm{d}\eta \nonumber\\
&=
\int_\eta \mathcal{L}(\eta) 
\int_v v \rho^v(t,v|\eta)
\int_w \rho^w(t,w|v,\eta)
\mathrm{d}w
\mathrm{d}v
\mathrm{d}\eta \nonumber\\
&=
\int_\eta \mathcal{L}(\eta) 
\int_v v \rho^v(t,v|\eta)
\mathrm{d}v
\mathrm{d}\eta,
\end{align}
where we use the normalization condition on the marginal density of $w$.
Furthermore, we integrate the general continuity equation (\ref{eq:CE-general}) with respect to $w$ and use \eqref{eq:pop-density}, yielding
\begin{equation}\label{eq:CE-2}
    \frac{\partial}{\partial t} 
    \rho^v(t,v|\eta)
    +
    \frac{\partial}{\partial v}
    \big[G^v(v,\langle w|v,\eta\rangle,s,\eta)\rho^v(t,v|\eta)\big]
    =0.
\end{equation}
To obtain this expression, we used the normalization condition on the marginal density of $w$ and the fact that the flux vanishes on the boundary $\partial w$. Finally, using the moment closure assumption (\ref{eq:mc1}), we obtain the resulting modified continuity equation,
\begin{equation}\label{eq:CE-cond}
   \frac{\partial}{\partial t} 
    \rho^v(t,v|\eta)
    +
    \frac{\partial}{\partial v}
    \big[G^v(v,\langle w|\eta\rangle,s,\eta)\rho^v(t,v|\eta)\big]
    =0.
\end{equation}
This modified continuity equation together with eq.~\ref{eq:MF-s}, an equation analogous to \eqref{eq:MF-w} for $\langle w|\eta\rangle$ and \eqref{eq:r-cond}-\eqref{eq:mean-v-cond} form a closed system for
the evolution of $\rho(t,v|\eta),\langle w|\eta\rangle$ and $s$.
Consideration of the steady state of the solution of this system yields
\begin{align}\label{eq:CE-sol}
\overline{\rho}^v(v|\eta) 
& \propto \frac{1}{G^v(v,\overline{\langle w|\eta\rangle},\overline{s},\eta)} \nonumber \\
& \propto \frac{1}{v(v-\alpha) - \overline{\langle w|\eta\rangle} + \eta + I_{\mathrm{ext}}+ g_{\mathrm{syn}}\overline{s}(e_r - v)},
\end{align}
where $\overline{\langle w|\eta\rangle}$ and $\overline{s}$ are the steady state
values of $\langle w|\eta\rangle$ and $s$, respectively.
%Since that the steady-state conditional probability, %$\overline{\rho}^v(v|\eta)$, 
%is inversely proportional to a quadratic function of $v$, it can be written as a Lorentzian distribution.
%\[ \overline{\rho}^v(v|\eta)=K\frac{\overline{x}(\eta)}{[v-\overline{y}(\eta)]^2+\overline{x}^2(\eta)}.\]

%===============================
\subsection{Lorentzian ansatz}
In this section, we will further simplify the expressions of the macroscopic variables $r(t)$ and $\langle v(t) \rangle$, and derive the mean-field approximation for the Izhikevich network by employing the Lorentzian ansatz \cite{Montbrio2015}.
To begin, we assume that the conditional probability $\rho^v(t,v|\eta)$ satisfies a time dependent version of \eqref{eq:CE-sol} and hence can be written in the form of Lorentzian distribution as follows,
\begin{equation}\label{eq:LA}
    \rho^v(t,v|\eta)=\frac{1}{\pi}\frac{x(t,\eta)}{\big [v-y(t,\eta)\big]^2+x^2(t,\eta)},
\end{equation}
where $x(t,\eta)$ and $y(t,\eta)$ are two time-dependent parameters defining half-width at half-maximum and location of the center, respectively. Moreover, $y(t,\eta)$ is  
%related to the mean membrane potential for each value of $\eta$ and 
defined via the Cauchy principal value as
\begin{equation}\label{eq:y}
y(t,\eta) = 
P.V. \int_v
v \rho^v(t,v|\eta)
\mathrm{d}v,
\end{equation}
the reason being that the Lorentz distribution only has a mean in principal value sense. So the mean membrane potential is related to $y(t,\eta)$ via
\begin{equation}\label{eq:mean-v-y}
\langle v(t) \rangle
=
\int_\eta
y(t,\eta)\mathcal{L}(\eta)
\mathrm{d}\eta.
\end{equation}
Under the condition
\begin{equation}
v_{\mathrm{peak}}
= -v_{\mathrm{reset}}
\rightarrow \infty    
\end{equation}
corresponding to $\theta = \pi$ in the theta model, the population firing rate defined as (\ref{eq:r-cond}) is also related to the Lorentzian coefficient through the intermediate expression,
\begin{equation}\label{eq:r_eta}
\begin{split}
r(\eta,t) &= 
\lim_{v\rightarrow v_{\mathrm{peak}}}
\rho^v(t,v|\eta)
G^v(v,\langle w|\eta \rangle,s,\eta) \\
& = \lim_{v_\mathrm{peak}\rightarrow \infty}\frac{1}{\pi}\frac{x(\eta,t)}{\big [v_\mathrm{peak}-y(\eta,t)\big]^2+x^2(\eta,t)}
\cdot \\
& \quad \quad \big [v_\mathrm{peak}(v_\mathrm{peak} - \alpha) 
 -\langle w|\eta\rangle + \eta  + I_\mathrm{ext} 
+ g_\mathrm{syn}s(e_r - v_\mathrm{peak})\big ] \\
& = \frac{1}{\pi}x(\eta,t).
\end{split}
\end{equation}
The total firing rate is then
\begin{equation}\label{eq:r-x}
r(t) 
=
\int_{\eta}{r(\eta,t)
\mathcal{L}(\eta)\mathrm{d}\eta}
=
\frac{1}{\pi}\int_\eta{x(\eta,t)
\mathcal{L}(\eta)\mathrm{d}\eta}.
\end{equation}

For the continuity equation (\ref{eq:CE-cond}),  we substitute the Lorentzian ansatz (\ref{eq:LA}) and equate the resulting equation in the powers of $v$, yielding
\begin{subequations}\label{eq:MF-xy}
\begin{align}
x' & = 2xy - (\alpha + g_{\mathrm{syn}}s)x, \label{eq:MF-x}\\
y' & = y(y - \alpha) - x^2 - \langle w|\eta \rangle + \eta + I_{\mathrm{ext}} + g_{\mathrm{syn}}s(e_r - y),\label{eq:MF-y}
\end{align}
\end{subequations}
where (\ref{eq:MF-x}) is from the coefficient of $v^2$ equal to zero; (\ref{eq:MF-y}) from the coefficient of $v$ equal to zero. Both of them lead to disappearance of the constant term. By defining a complex variable $z(t,\eta) = x(t,\eta)+iy(t,\eta)$, we write  (\ref{eq:MF-xy}) in complex form as
\begin{equation}\label{eq:MF-complex}
\frac{\partial}{\partial t}
z(t,\eta) = 
i \Big[
- z^2(t,\eta)
+
iz(t,\eta)(\alpha + g_{\mathrm{syn}}s)
-\langle w|\eta \rangle + \eta  + I_{\mathrm{ext}} +g_{\mathrm{syn}}s e_r
\Big].
\end{equation}
At this point, we have obtained the mean-field approximation (\ref{eq:MF-xy}) or (\ref{eq:MF-complex}) for the Izhikevich network. It could be used to determine the two macroscopic variables $\langle v \rangle$ and $r$ via (\ref{eq:mean-v-y}) and (\ref{eq:r-x}). However, the evolution of \eqref{eq:MF-complex} depends on the heterogeneous current $\eta$ and $\langle w|\eta
\rangle$ and hence on the distribution $\mathcal{L}(\eta)$.

%===============================
\subsection{Heterogeneity with Lorentzian distribution}
Further derivation of the mean-field description in terms of the macroscopic observables $r$ and $\langle v \rangle$ depends on the distribution of the heterogeneous parameter, $\eta$.  
Specifically, we choose the heterogeneous current $\eta$ to have a Lorentzian distribution with center $\eta$ and  half-width at half-maximum $\Delta_\eta$, i.e.,
\begin{equation}\label{eq:LA-para}
\mathcal{L}(\eta) 
=
\frac{1}{\pi}
\frac{\Delta_\eta}{(\eta - \bar \eta)^2 + \Delta_\eta^2}
\end{equation}
Then, we apply the residue theorem to compute the integrals in (\ref{eq:mean-v-y}) and (\ref{eq:r-x}) for $\eta \in (-\infty, \infty)$, to obtain
\begin{subequations}
\label{eq:rv-xy}
\begin{align}
r(t) &= 
\frac{1}{\pi}x(t,\bar \eta-i\Delta_\eta), \\
\langle v(t) \rangle &=
y(t,\bar \eta-i\Delta_\eta).
\end{align}
\end{subequations}
Further, considering $\pi r(t) + i\langle v(t) \rangle = z(\bar \eta - i\Delta_\eta)$, evaluating the complex equation (\ref{eq:MF-complex}) at $\eta = \bar \eta-i\Delta_\eta$ and taking into account the formula that
\begin{equation}
\langle w \rangle = 
\int_\eta
\langle w|\eta \rangle
\mathcal{L}(\eta)
\mathrm{d}\eta
\end{equation}
yields the mean-field system of the firing rate equations (FREs) given by
\begin{subequations}\label{eq:MF-rv}
\begin{align*}
r'
& = 
\Delta_\eta/\pi 
+
2r \langle v \rangle
-
\big(\alpha + g_{\mathrm{syn}}s \big)r,
\\
\langle v \rangle'
& =
\langle v \rangle^2
-
\alpha  \langle v \rangle
-
\langle w \rangle
+
\bar \eta + I_{\mathrm{ext}}
+
g_{\mathrm{syn}} s 
\big(
e_r - \langle v \rangle
\big)
-
\pi^2 r^2.
\end{align*}
\end{subequations}
Note that the distribution  $\mathcal{L}(\eta)$ can be arbitrary. Particularly, if $\mathcal{L}(\eta)$ has $n$ poles in the lower half $\eta$-plane, one can readily obtain $n$ sets of complex-valued mean-field ODEs by evaluating the integrals (\ref{eq:mean-v-y}) and (\ref{eq:r-x}) \cite{Ott2008}.  Lorentzian distribution is a mere mathematical convenience since it has only one pole as required.

Recalling that we already obtained the dynamical system for the mean adaptation current (\ref{eq:MF-w}) and synapses (\ref{eq:MF-s}), we obtain the reduction of the network of Izhikevich neurons (\ref{eq:network-Izh})-(\ref{eq:single-exp-synapse}) to the following 
the mean-field system of ODEs,
\begin{subequations}\label{eq:mf-izh}
\begin{align}
r'
& = 
\Delta_{\eta}/\pi 
+
2r \langle v \rangle
-
\big(\alpha + g_{\mathrm{syn}}s \big)r,
\\
\langle v \rangle'
& =
\langle v \rangle^2
- \alpha \langle v \rangle
-
\langle w \rangle
+
\bar \eta + I_{\mathrm{ext}}
+
g_{\mathrm{syn}} s 
\big(
e_r - \langle v \rangle
\big)
-
\pi^2 r^2
\\
\langle w \rangle'
&=
a
\left (
b 
\langle v \rangle
-
\langle w \rangle
\right )
+
w_{\mathrm{jump}} r
\\
s'
&=
-s/\tau_s
+
s_{\mathrm{jump}} r
\end{align}
\end{subequations}

%%%%%%%%%%%%%%%%%%%%%%%%%%%%%%%%%%%%%

\section{NUMERICAL ANALYSIS}\label{sec:numerics}
We now numerically examine the dynamics of the mean-field model and demonstrate its validity in terms of reproducing the macroscopic dynamics of the network of Izhikevich neurons.

We consider an all-to-all coupled network with synapses governed by the single exponential model. The parameter values used in all simulations can be found in Table~\ref{tab:para-dimensionless}, unless otherwise specified in a figure. The corresponding dimensional form of the network model and parameter values are given in \ref{app:nondim}. Most of these values are taken from \cite{Nicola2013bif,Nicola2013hetero} which were originally fit by \cite{Dur-e-Ahmad2012} to hippocampal CA3 pyramidal neuron data from \cite{Hemond2008}. The exceptions are  $v_{\mathrm{peak}}$ and $v_{\mathrm{reset}}$ which are, respectively, set to large positive and negative numbers. This is to approximate $v_{\mathrm{peak}}\rightarrow\infty$ and $v_{\mathrm{reset}}\rightarrow -\infty$ which is required for the QIF model to be well-approximated by the theta model \cite{Izhikevich2007}. 
Numerical simulation was done by using the Euler's method in MATLAB, with time step $dt = 10^{-3}$ and numerical continuation by using the software XPPAUT \cite{Ermentrout2002}. 

\begin{table}[ht!]
\centering\footnotesize
\begin{threeparttable}
\caption{Dimensionless parameters for the network of Izhikevich neurons}
\label{tab:para-dimensionless}
\begin{tabular}{l l l l}
  \toprule
  Parameter & Value & Parameter & Value\\
  \midrule
  $\alpha$
  & $0.6215$
  & $\tau_s$
  & $2.6$ \\
  $g_{\mathrm{syn}}$
  & $1.2308$
  & $e_r$
  & $1$ \\
  $a$
  & $0.0077$ 
  & $b$
  & $-0.0062$ \\
  $s_{\mathrm{jump}}$
  & $1.2308$ 
  & $w_{\mathrm{jump}}$
  & $0.0189$  \\
  $v_{\mathrm{peak}}$
  & 200
  & $v_{\mathrm{reset}}$
  & -200 \\
\bottomrule
\end{tabular}
\begin{tablenotes}[para,flushleft]
Note: These parameters apply unless otherwise indicated.
\end{tablenotes}
\end{threeparttable}
\end{table}

\begin{figure}[ht!]
\centering
\includegraphics[width=\figwidth]{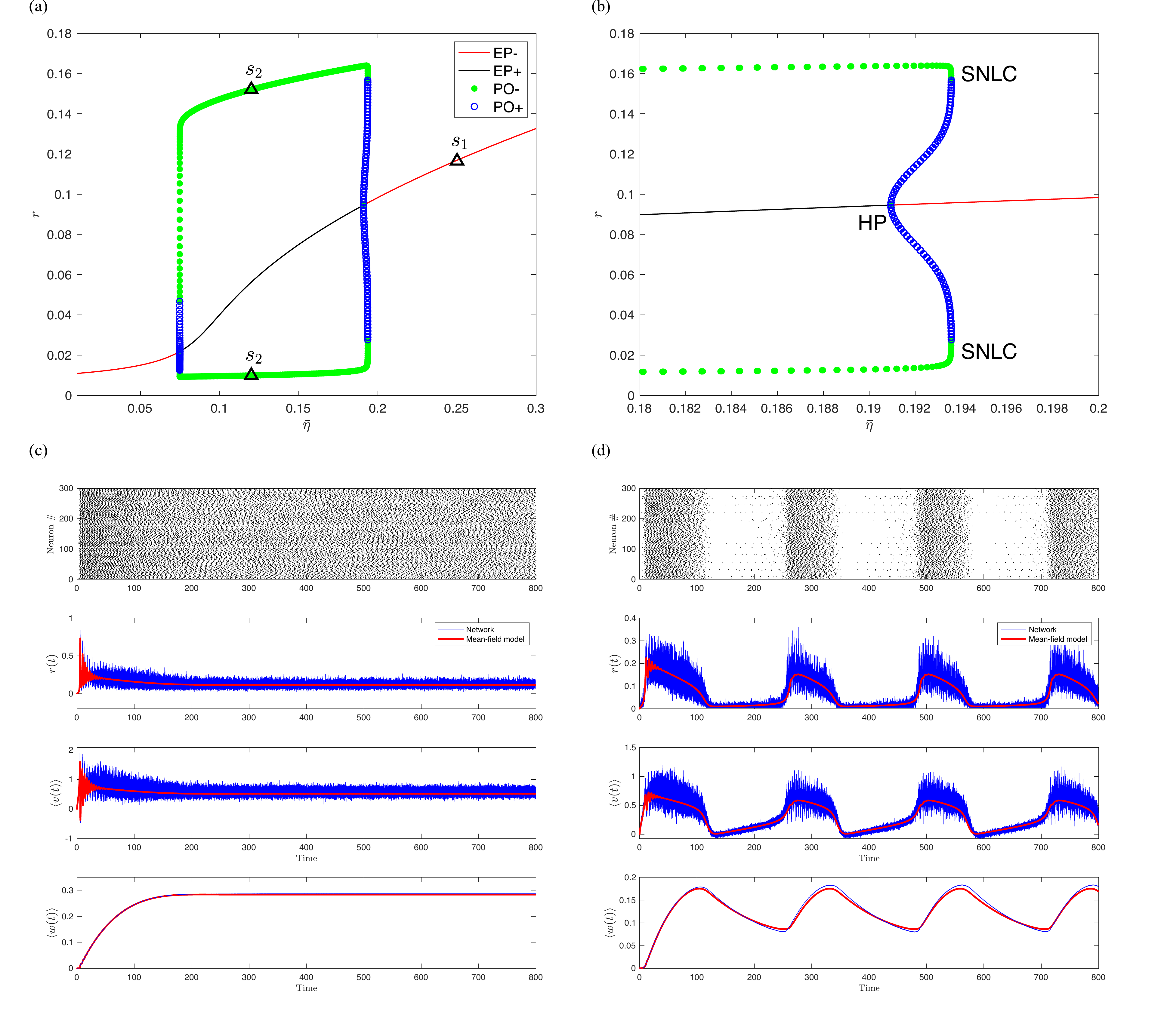}
\caption{Comparison of one-parameter bifurcation diagram and time evolution. 
(a): Bifurcation diagram of the mean-field model (\ref{eq:mf-izh}) with respect to the mean intrinsic current $\bar \eta$. The red (black) lines  correspond to stable (unstable) equilibrium points and green (blue) dots correspond to stable (unstable) limit cycles.
(b): Blow-up of the bifurcation diagram near $\bar \eta = 0.19$. Bistability is induced by a subcritical Hopf bifurcation (HB) and saddle-node bifurcations of limit cycles (SNLC). Similar qualitative changes occur at  $\bar \eta approx 0.07$. 
(c) \& (d): Comparison of the temporal behaviour of  the network of Izhikevich neurons (\ref{eq:network-Izh})-(\ref{eq:single-exp-synapse}) and the corresponding mean-field model (\ref{eq:mf-izh})  when $\bar \eta = 0.25$ ($s_1$ in (a)) and $\bar \eta  = 0.12$ ($s_2$ in (a)). First row of panels are the raster plots of $300$ randomly selected Izhikevich neurons from the $N = 10^4$ in the population. Other rows of panels show, respectively, the population firing rate $r(t)$, mean membrane potential $\langle v(t) \rangle$ and mean adaptation variable $\langle w(t) \rangle$ obtained from simulations of the full network (blue) and the mean-field model (red). Parameters: $\Delta_\eta = 0.02$, $I_{\mathrm{ext}}=0$, others are as given in Table~\ref{tab:para-dimensionless}.}
\label{fig:1pop-merge1}
\end{figure}

 We begin with the bifurcation analysis of the mean-field model (\ref{eq:mf-izh}). Fig.~\ref{fig:1pop-merge1} (a) and its blow-up (b) show how the population firing rate $r$ qualitatively changes as the mean intrinsic current $\bar \eta$ is varied. There are two subcritical Andronov-Hopf bifurcations (HP) at  $\bar \eta = \bar \eta_{HP} \approx 0.191$ and $0.07$, respectively. Unstable limit cycles emerge from these bifurcations and collide with the stable limit cycles in a saddle-node bifurcation of limit cycles (SNLC) for some $\bar \eta = \bar \eta_{SNLC} > \bar \eta_{HP}$ (right branches) or $ < \bar \eta_{HP}$ (left branches). The system displays two small ranges of bistability between the Hopf and SNLC bifurcations. The stable limit cycles (green dots) correspond to bursting solutions in the full network and stable equilibrium points (red lines) correspond to the tonic firing. This is clearly reflected in the time series of macroscopic variables $r(t)$, $\langle v(t) \rangle$ and $\langle w(t) \rangle$ in Fig.~\ref{fig:1pop-merge1} (c) and (d). The mean-field equations (\ref{eq:mf-izh}) exactly predict the behaviour of the full network, including the damped oscillations in (c) and the frequency of stable oscillations in (d). %at the macroscopic level. \cite{Montbrio2015}.

% matlab codes from "1_IzhikevichNeuron_Heter_Lorentzian_burst***", results and XPPAUT codes see "Paper_MF_Izhikevich_Neuron/OnePopulation"

%These results are largely consistent with prior work on homogeneous  and heterogeneous \cite{Nicola2013hetero} networks of Izhikevich neurons. 
Prior work has shown that population bursting in the networks of Izhikevich neurons is due to a balance between the inputs (intrinsic and external applied currents and synaptic inputs), which cause the neurons to spike, and the slow adaptation current, which can terminate spiking \cite{Dur-e-Ahmad2012,Nicola2013bif,Nicola2013hetero}
. For a given level of adaptation there must be sufficient input, but not too much. Hence the bursting in  Fig.~\ref{fig:1pop-merge1}(a) occurs when the mean intrinsic current $\bar{\eta}$ is not too small and not too large.
%Our results improve on those of \cite{Nicola2013bif,Nicola2013hetero} in the sense that the mean-field model gives a very accurate representation of the frequency of bursting.
%
Note in Fig.~\ref{fig:1pop-merge1}(d) that even when the population is bursting, a small subset of neurons in the population do not burst but remain tonically firing. This is due to the distribution of the heterogeneous input current. A small number of neurons will receive a large enough input current that the adaptation is not strong enough to cause the neuron to burst.

\begin{figure}[ht!]
\centering
\includegraphics[width=\figwidth]{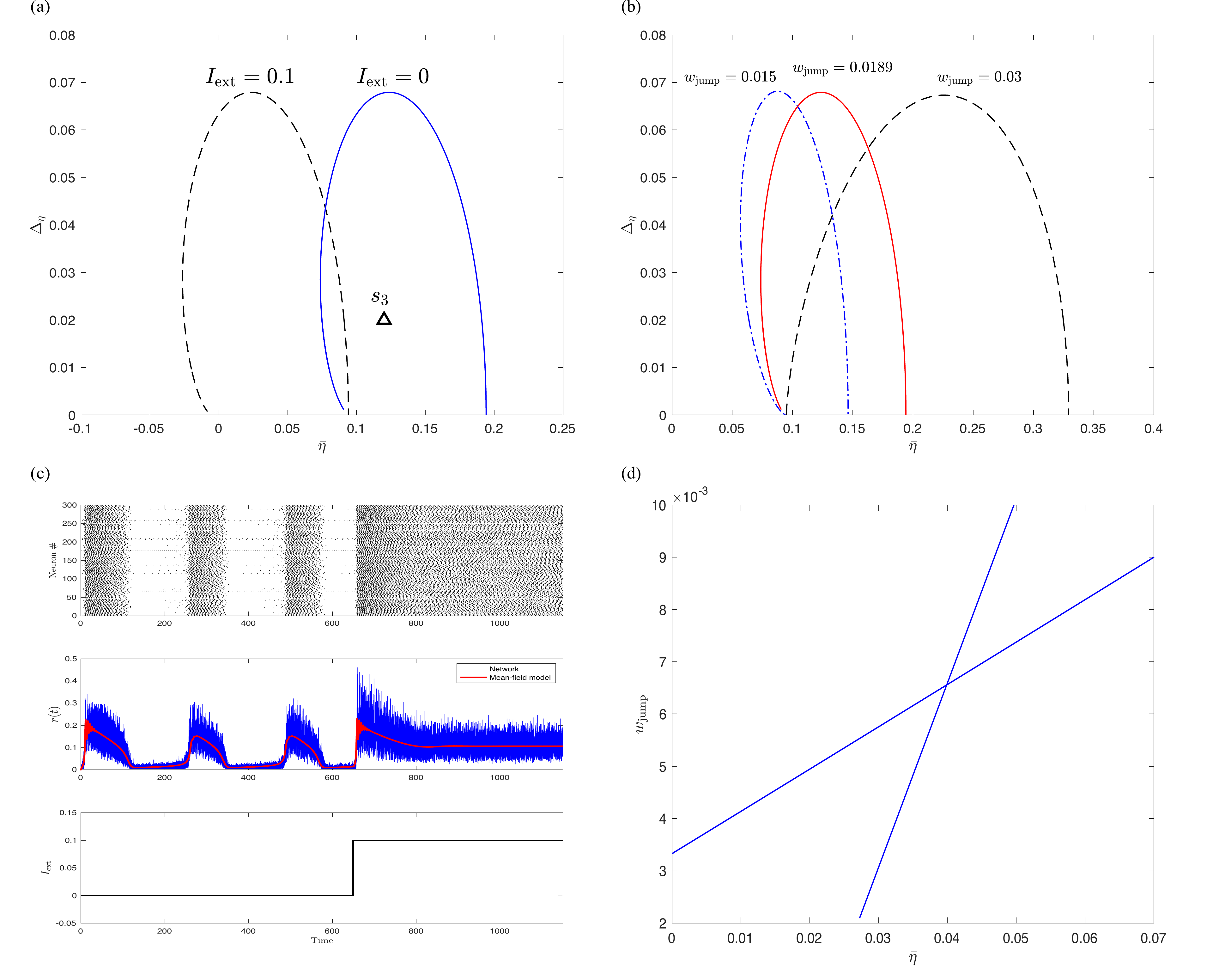}
\caption{Two-parameter bifurcation diagrams and associated time evolution. 
(a) \& (b): Hopf bifurcation manifolds for the mean-field model (\ref{eq:mf-izh}) in the parameter space of the mean and half-width of the distribution of intrinsic applied current.  %associated with the transition between bursting and tonic firing.  
(a) Two values of the external current $I_{\mathrm{ext}}$ when $w_{\mathrm{jump}} = 0.0189$. (b) Three values of the adaptation jump parameter, $w_{\mathrm{jump}}$, when $I_{\mathrm{ext}}=0$. 
(c): Dynamics of the network of Izhikevich neurons (\ref{eq:network-Izh})-(\ref{eq:single-exp-synapse}) compared with the corresponding mean-field model (\ref{eq:mf-izh}) when $\bar \eta = 0.12$, $\Delta_\eta = 0.02$ ($s_3$ in (a)). First panel is the raster plot of the spiking activity. The instantaneous population firing rate of the network and the mean-field model are depicted in blue and red, respectively. Stimulus $I_\mathrm{ext}(t)$ is shown in the last panel. At time $t=650$, a current $I_\mathrm{ext}=0.1$ is applied to all neurons.
(d): Double Hopf bifurcation induced by the intersection of two Hopf bifurcations in the   $(\bar \eta, w_\mathrm{jump})$ parameter space when $\Delta_\eta = 0.02$. 
Parameters: $N = 10^4$. Others are given in Table~\ref{tab:para-dimensionless}.}
\label{fig:1pop-merge2}
\end{figure}

We can also determine the Andronov-Hopf bifurcation manifolds in the $(\bar \eta, \Delta_\eta)$ parameter space for the mean-field model as shown in Fig.~\ref{fig:1pop-merge2} (a) and (b). These curves are associated with the transition between bursting (below the curves) and tonic firing (above the curves). We see from the network raster plots and time series of $r(t)$ [Fig.~\ref{fig:1pop-merge2} (c)] that the rhythmic regime disappears if the external drive $I_\mathrm{ext}$ is sufficiently strong. Additionally, Fig.~\ref{fig:1pop-merge2} (d) shows a Hopf-Hopf bifurcation resulting from intersection of two Andronov-Hopf bifurcations in the $(\bar \eta, w_\mathrm{jump})$ parameter space. The two curves look like straight lines in such narrow scales.
Secondary bifurcations can emanate from this co-dimension two bifurcation point, leading to, for example, quasiperiodic behaviour \cite{Guck_Holmes}.
%that and imply a local birth of "chaos". 
We leave further investigation of this bifurcation for future work.
% matlab codes from "1_IzhikevichNeuron_Heter_Lorentzian_Change_I***", results and XPPAUT codes see "Paper_MF_Izhikevich_Neuron/OnePopulation"

%%%%%%%%%%%%%%%%%%%%%%%%%%%%%%%%%%%%%

\section{EXTENSION TO TWO-COUPLED POPULATIONS}\label{sec:2pop}
A large-scale neural network can also be regarded as several coupled groups by considering different properties of cells in the network. For example, neurons are grouped into excitatory and inhibitory populations based on the type of synapses they form, e.g., \cite{WilsonCowan1972,Dumont2019}; or into strongly and weakly adapting populations based on the amount of spike frequency adaptation they exhibit, e.g., \cite{Nicola2013bif,Hemond2008}. 
In the section, we consider a network of strongly adapting neurons (population $p$) and weakly adapting neurons (population $q$) all-to-all connected with single exponential synapses. Each neuron is characterized by the Izhikevich model given by 
\begin{subequations}\label{eq:2pop-network-izh}
\begin{align}
    v'_{m,k} 
    &=
    v_{m,k}(v_{m,k} - \alpha_m) 
    -
    w_{m,k} 
    +
    \eta_{m,k} + I^{\mathrm{ext}}_{m}
    + 
    I^{\mathrm{syn}}_{m,k}, \\
    w'_{m,k}
    & = 
    a_m(b_m v_{m,k} - w_{m,k}), \\
    \mathrm{if} \; v_{m,k} 
    &
    \geq v^{\mathrm{peak}}_m, 
    \mathrm{then} \;
    v_{m,k} \leftarrow v^{\mathrm{reset}}_m\;
    \mathrm{and}\;
    w_{m,k} \leftarrow w_{m,k} + w^{\mathrm{jump}}_m,
\end{align}
\end{subequations}
where $m=p,q$ represents the two populations with $N_p$ and $N_q$ cells, respectively. The subscript $\{m,k\}$ denotes the $k$th neuron in population $m$. The subscript with only $\{m\}$ represents the corresponding parameter is homogeneous within the population $m$, but heterogeneous across the two populations. 
The total synaptic current $I^{\mathrm{syn}}_{m,k}$ depends on the cell type. We require two maximal synaptic conductances, $g^\mathrm{syn}_{p,p}$ and $g^\mathrm{syn}_{q,q}$, within the populations and two, $g^\mathrm{syn}_{p,q}$ and $g^\mathrm{syn}_{q,p}$, between the populations. Then, we have
\begin{subequations}\label{eq:2pop-net-Isyn}
\begin{align}
    I^{\mathrm{syn}}_{p,k}(t)
    &=
    \big [
    \kappa g^{\mathrm{syn}}_{p,p}s_p 
    +
    (1-\kappa) g^{\mathrm{syn}}_{p,q}s_q 
    \big ]
    \left( e^r_p - v_{p,k} \right)
    \equiv G_pS_p \left( e^r_p - v_{p,k} \right)
    \\
    I^{\mathrm{syn}}_{q,k}(t)
    &=
    \big [
    \kappa g^{\mathrm{syn}}_{q,p}s_p 
    +
    (1-\kappa) g^{\mathrm{syn}}_{q,q}s_q
    \big ]
    \left( e^r_q - v_{q,k} \right)
    \equiv G_qS_q \left( e^r_q - v_{q,k} \right),
\end{align}
\end{subequations}
where $\kappa = \frac{N_p}{N_p + N_q}$ is the proportion of strongly adapting neurons in the network and $s_p$ (respectively, $s_q$) represents the proportion of open synapses due to neurons in the strongly (respectively, weakly) adapting population. These gating variables are governed by the single exponential synapse model,
\begin{equation}\label{eq:2pop-net-s}
    s'_{m} 
    =
    -
    \frac{s_m}{\tau^s_m} 
    +
    \frac{s_m^{\mathrm{jump}}}{N_m}
    \sum_{k=1}^{N_m}
    \sum_{t_{m,k}^j<t}
    \delta(t-t_{m,k}^j), \quad m=p,\;q.
\end{equation}
Thus, we can apply the Lorentzian ansatz and the method in the previous sections by considering the two populations to be described by their own distinct density functions,
\begin{equation*}
\rho_m(t,v_m,w_m,\eta_m)=\rho^w_m(t,w_m|v_m,\eta_m)\rho^v_m(t,v_m|\eta_m)\mathcal{L}_m(\eta_m), \; m=p,q.
\end{equation*}
% details see appendix of draft 3 
%
Finally, assuming the Lorentzian distribution of the heterogeneous currents for both populations,
\begin{equation}\label{eq:2pop-LA}
    \mathcal{L}(\eta_m) 
    =
    \frac{1}{\pi}
    \frac{\Delta^\eta_m}{(\eta_m - \bar \eta_m)^2 + (\Delta^\eta_m)^2}, \quad m=p,\;q,
\end{equation}
We obtain the mean-field system consisting of a set of eight differential equations. Three differential equations describe the mean-field quantities for each population,
\begin{subequations}\label{eq:2pop-mf-izh-p}
\begin{align}
r_p' 
& =
\Delta^\eta_p/\pi 
+ 2r_p \langle v \rangle_p
- r_p [G_p S_p+\alpha_p ] 
\\
\langle v \rangle_p '
& =
\langle v \rangle_p^2
- \alpha_p \langle v \rangle_p
- \langle w\rangle_p 
+ \bar \eta_p + I^{\mathrm{ext}}_p 
+ G_p S_p [e^r_p- \langle v \rangle_p ]
- \pi^2r_p^2
\\
\langle w\rangle'_p
& =
a_p
[
b_p \langle v \rangle_p
- \langle w \rangle_p
]
+ w_p^{\mathrm{jump}} r_p
\end{align}
\end{subequations}
for the population $p$ with strong adaptation and
\begin{subequations}\label{eq:2pop-mf-izh-q}
\begin{align}
r_q' 
& =
\Delta^\eta_q/\pi 
+ 2r_q \langle v \rangle_q
- r_q [G_q S_q+\alpha_q ] 
\\
\langle v \rangle_q '
& =
\langle v \rangle_q^2
- \alpha_q \langle v \rangle_q
- \langle w\rangle_q 
+ \bar \eta_q + I^{\mathrm{ext}}_q 
+ G_q S_q [e^r_q- \langle v \rangle_q ]
- \pi^2r_q^2
\\
\langle w\rangle'_q
& =
a_q
[
b_q \langle v \rangle_q
- \langle w \rangle_q
]
+ w_q^{\mathrm{jump}} r_q
\end{align}
\end{subequations}
for the population $q$ with weak adaptation.
These two subsystems are coupled through synaptic currents as given in eq. (\ref{eq:2pop-net-Isyn}), with the synaptic dynamics governed by
\begin{subequations}\label{eq:2pop-mf-s}
\begin{align}
s'_{p} 
&=
- s_p / \tau^s_p
+ s_p^{\mathrm{jump}} r_p 
\\
s'_{q} 
&=
- s_q / \tau^s_q
+ s_q^{\mathrm{jump}} r_q.
\end{align}
\end{subequations}

In the following we analyze the dynamics of the mean-field model and examine how well it reproduces the macroscopic activities of the two-population network of Izhikevich neurons. The parameter values can be found in Table~\ref{tab:2pop-para-dimensionless}. The only parameters that differ between the two populations are those that govern the adaptation levels, i.e., the after-spike jump sizes $w^\mathrm{jump}_m$ and time constants $a_m$, where $m=p,q$.

\begin{table}[ht!]
\footnotesize 
\centering
\begin{threeparttable}
\caption{Dimensionless parameters for the two coupled Izhikevich network}
\label{tab:2pop-para-dimensionless}
\begin{tabular}{l l l l}
  \toprule
  Parameter & Value & Parameter & Value\\
  \midrule
  $w^{\mathrm{jump}}_p$
  & $0.0189$
  & $w^{\mathrm{jump}}_q$
  & $0.0095$ \\
  %%%%%%
  $a_p$
  & $0.0077$
  & $a_q$
  & $0.077$ \\
  %%%%%%
  $g^{\mathrm{syn}}_{p,p}$
  & $1.2308$
  & $g^{\mathrm{syn}}_{q,q}$
  & $1.2308$ \\
  %%%%%%
  $g^{\mathrm{syn}}_{p,q}$
  & $1.2308$
  & $g^{\mathrm{syn}}_{q,p}$
  & $1.2308$ \\
  %%%%%%
  $I^\mathrm{ext}_p$
  & $0$
  & $I^\mathrm{ext}_q$
  & $0$ \\
  %%%%%%
  $N_p$
  & $8000$
  & $N_q$
  & $2000$ \\
\bottomrule
\end{tabular}
\begin{tablenotes}[para,flushleft]
Note: All parameters that are not mentioned here can be found in Table~\ref{tab:para-dimensionless}. These parameters apply unless otherwise indicated.
\end{tablenotes}
\end{threeparttable}
\end{table}

\begin{figure}[ht!]
\centering
\includegraphics[width=\figwidth]{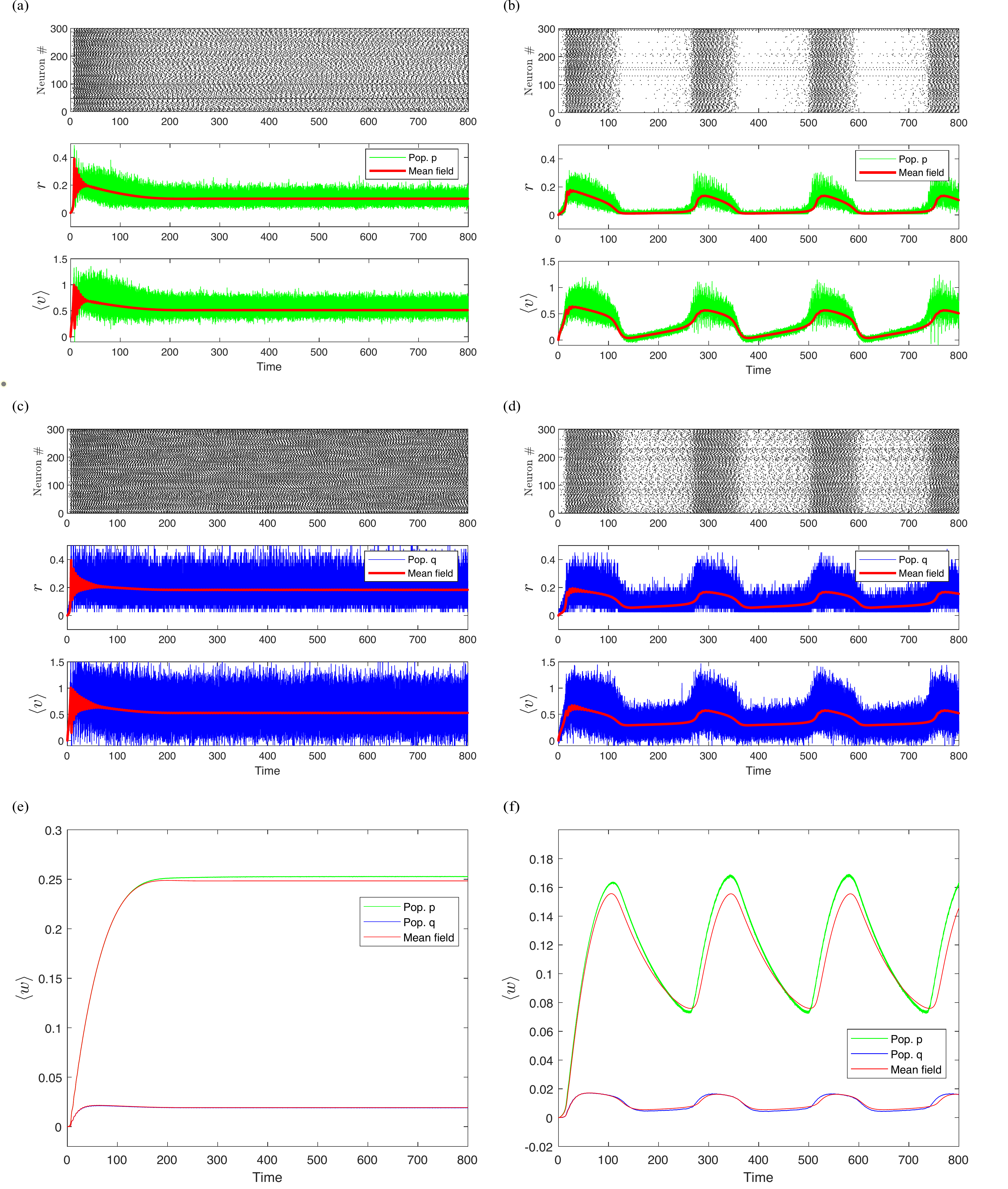}
\caption{Time evolution comparison between the network of Izhikevich neurons (\ref{eq:2pop-network-izh})-(\ref{eq:2pop-net-s}) and the corresponding mean-field model (\ref{eq:2pop-mf-izh-p})-(\ref{eq:2pop-mf-s}).
The network consists of 8000 strongly adapting neurons (population $p$) and 2000 weakly adapting neurons (population $q$).
The left column of panels shows behaviour of two populations when $\bar \eta_p = \bar \eta_q = 0.18$ ($s_1$ in the bifurcation diagram Fig. \ref{fig:2pop-merge1}a).
The right column is the case when $\bar \eta_p = \bar \eta_q = 0.08$ ($s_2$ in Fig. \ref{fig:2pop-merge1}a).
(a) \& (b) correspond to the population $p$, (c) \& (d) the population $q$. The first rows in panels (a-d) are the raster plots of the spiking activity.
Shown in green (population $p$) and blue (population $q$) are the population firing rates, mean membrane potentials and mean adaptation variables (e \& f) obtained from the full network. Shown in red are the corresponding variables in the mean-field models.
Parameters: $\Delta^{\eta}_p = \Delta^{\eta}_q = 0.02$, others are shown in Table~\ref{tab:2pop-para-dimensionless}.}
\label{fig:2pop-TimeEvolution}
\end{figure}

Fig.~\ref{fig:2pop-TimeEvolution} shows that the dynamics of the two-population network are exactly described by the reduced mean-field description in the tonic firing (equilibrium points) and bursting (periodic orbits) regimes. Note that in the rhythmic regime, a relatively larger fraction of neurons in the weakly adapting population is tonically firing [see (d)]. This makes sense since the bursting is due to the balance of inputs and adaptation. In the weakly adapting population a larger fraction of neurons receive sufficient input to prevent them from bursting.

% matlab codes from "2_Izhikevich_StrongWeakAdapt_Heter_Lorentzian_DifPara***", results and XPPAUT codes see "Paper_MF_Izhikevich_Neuron/SA-WA_SameParaAsOnePop"

%============================

\begin{figure}[ht!]
\centering
\includegraphics[width=\textwidth]{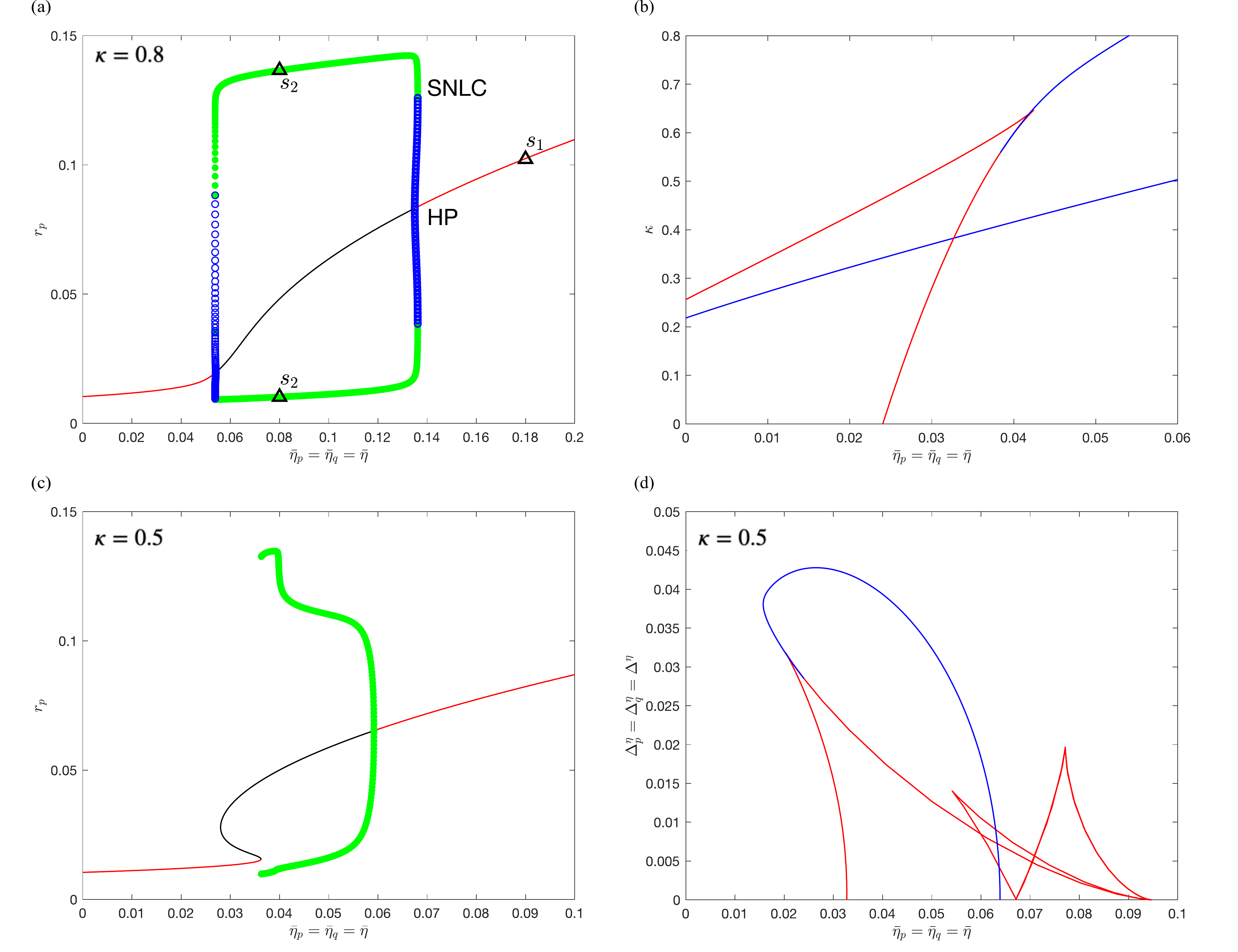}
\caption{Bifurcation diagram of the mean-field model (\ref{eq:2pop-mf-izh-p})-(\ref{eq:2pop-mf-s}).
(a): Qualitative changes of the population firing rate $r_p$ with respect to the mean intrinsic current $\bar \eta_p = \bar \eta_q = \bar \eta$ when the proportion of strong adapting neurons $\kappa = 0.8$. The red (black) lines  correspond to the stable (unstable) equilibrium points, and green (blue) dots correspond to stable (unstable) limit cycles. 
(b): Bifurcation curves plotted in the $(\bar \eta, \kappa)$-parameter plane. Shown in red are the saddle-node bifurcations. Shown in blue are the Hopf bifurcations.
(c): Qualitative changes of the population firing rate $r_p$ with respect to $\bar \eta_p = \bar \eta_q = \bar \eta$ when $\kappa = 0.5$. A supercritical Hopf bifurcation occurs at $\bar \eta_p = \bar \eta_q \approx 0.06$ and two saddle node bifurcations at $\bar \eta_p = \bar \eta_q \approx 0.028$, $0.036$, respectively.
(d): Bifurcation curves plotted in the $(\bar \eta, \Delta^\eta)$-parameter plane when $\kappa = 0.5$. Shown in red are the saddle-node bifurcations. Shown in blue is the Hopf bifurcation. 
Parameters: $\Delta^\eta_p = \Delta^\eta_q = 0.02$, $N_p + N_q = 10^4$  and others are shown in Table~\ref{tab:2pop-para-dimensionless}.}
\label{fig:2pop-merge1}
\end{figure}
% matlab codes from "2_Izhikevich_StrongWeakAdapt_Heter_Lorentzian_DifPara***", XPPAUT and other results see "Paper_MF_Izhikevich_Neuron"/ SA-WA_SameParaAsOnePop

Additionally, the mean-field model for the network of two coupled Izhikevich populations  involves more complicated bifurcations compared with that of the single-population network of strongly adapting Izhikevich neurons studied in the previous section.
Bifurcation analysis reveals that when the proportion of strong adapting neurons is $\kappa = 0.8$, the sequence of bifurcation is largely the same as when there is a single population of strongly adapting neurons (compare Fig.~\ref{fig:2pop-merge1}(a) with  Fig.~\ref{fig:1pop-merge1}(a)). With this value of $\kappa$, as the mean intrinsic current is increased stable periodic behaviour in the mean-field system (\ref{eq:2pop-mf-izh-p})-(\ref{eq:2pop-mf-s}) is initiated by a saddle-node bifurcation of limit cycles connected to a subcritical Andronov-Hopf bifurcation at $\bar \eta_p = \bar \eta_q  \approx 0.05$ and terminated by the same
sequence in reverse at $\bar \eta_p = \bar \eta_q \approx 0.14$.
%[see Fig.~\ref{fig:2pop-merge1} (a)]. 
The system displays bistability in the narrow regions
between the SNLC and HP. 
%around $\bar \eta_p = \bar \eta_q = 0.14$ and $\bar \eta_p = \bar \eta_q  = 0.05$, respectively.
%
However, complex bifurcations occur when changing the proportion of strongly adapting neurons. Fig.~\ref{fig:2pop-merge1} (b) show bifurcation curves plotted in the parameter planes of $(\bar \eta, \kappa)$. Saddle-node bifurcation curves (red) meet and form cusp points or tangentially intersect the curves of Andronov-Hopf bifurcation (blue) and produce the zero-Hopf bifurcation. When $\kappa = 0.8$, the system has two Hopf points, as shown in Fig.~\ref{fig:2pop-merge1} (a). When the proportion is reduced to $\kappa = 0.5$, the system undergoes two saddle-node bifurcation and one Andronov-Hopf bifurcation. These behaviours are also shown in the one-parameter bifurcation diagram Fig.~\ref{fig:2pop-merge1} (c).
%Fig.~\ref{fig:2pop-merge1} (c) shows that when the proportion is reduced to $\kappa = 0.5$, two saddle node of bifurcations appear. 
%
Further, one can see from Fig.~\ref{fig:2pop-merge1} (c) that
the stable period behaviour is now initiated by what appears to be a saddle-node on an invariant circle bifurcation or a homoclinic bifurcation (at $\bar \eta_p = \bar \eta_q \approx 0.036$) and terminated by  
a supercritical Andronov-Hopf bifurcation at  $\bar \eta_p = \bar \eta_q \approx 0.06$. 
Finally, Fig.~\ref{fig:2pop-merge1} (d)  shows a two-parameter bifurcation diagram in $(\bar \eta, \Delta^\eta)$ when $\kappa$ stays at $0.5$.  Complicated bifurcation structures, including Bogdanov-Takens bifurcation, may occur for nearby parameter values. We leave further investigation to future work.

%============================

\begin{figure}[ht!]
\centering
\includegraphics[width=\textwidth]{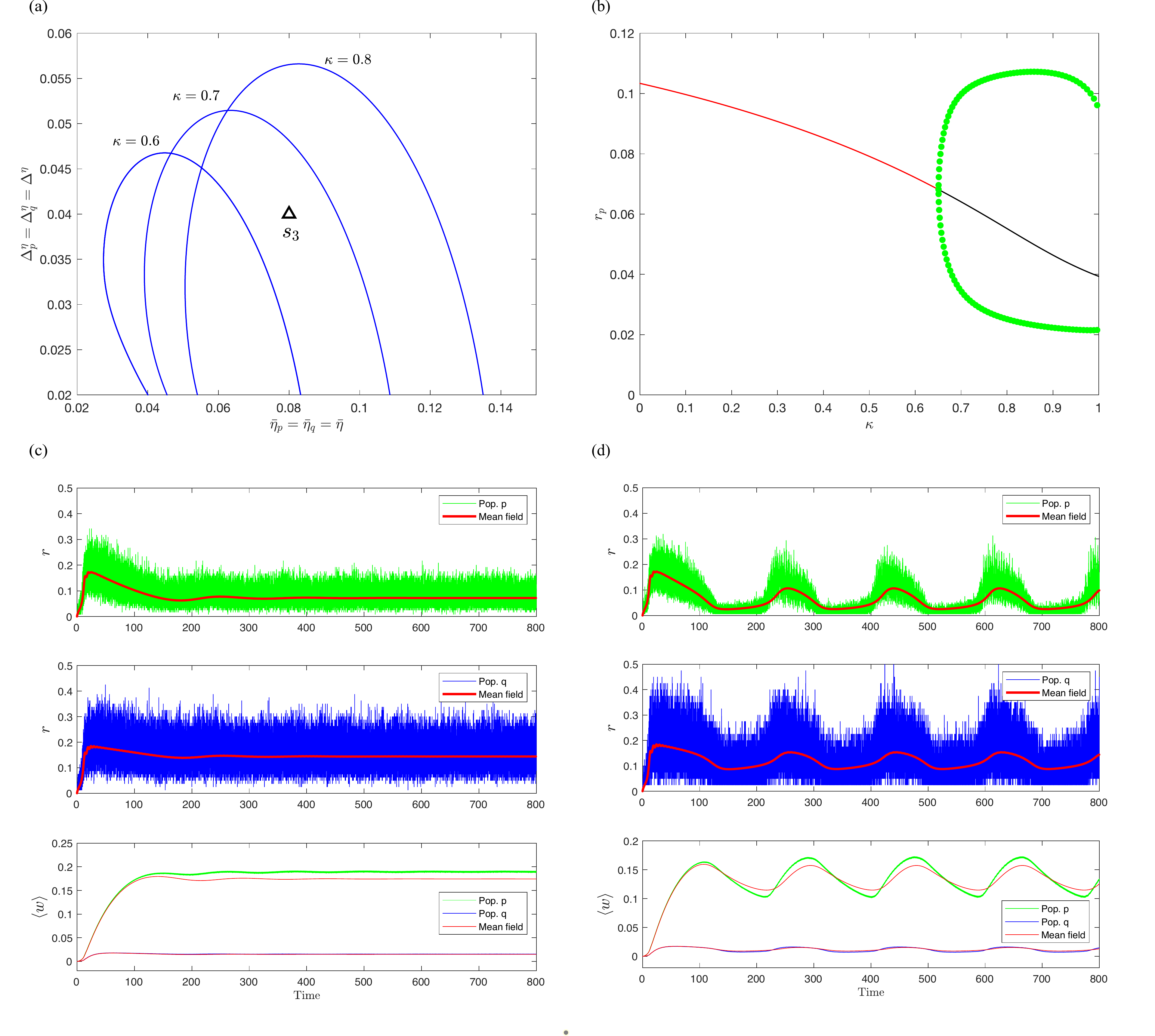}
\caption{Bifurcation diagram and time evolution comparison.
(a): Hopf bifurcation manifolds for the mean-field model (\ref{eq:2pop-mf-izh-p})-(\ref{eq:2pop-mf-s}) with different values of proportion of strongly adapting neurons $\kappa$.
(b): Bifurcation diagram of the population firing rate $r_p$  with respect to $\kappa$. The red (black) lines  correspond to the stable (unstable) equilibrium points, and green dots correspond to stable limit cycles.
(c) \& (d): Time evolution comparison between the full network (\ref{eq:2pop-network-izh})-(\ref{eq:2pop-net-s}) and the corresponding mean-field model (\ref{eq:2pop-mf-izh-p})-(\ref{eq:2pop-mf-s}) when $\kappa = 0.6$ and $0.8$, respectively. 
Shown in green (population $p$)  and blue (population $q$) are the population firing rates and mean adaptation variables obtained from the full network. Shown in red are the corresponding variables in the mean-field models.
The parameter values are shown in Table~\ref{tab:2pop-para-dimensionless}, including $\bar \eta_p = \bar \eta_q = 0.08$ and $\Delta^{\eta}_p = \Delta^{\eta}_q = 0.04$ corresponding to $s_3$ in (a).
}
\label{fig:2pop-merge3}
\end{figure}
% matlab codes from "2_Izhikevich_StrongWeakAdapt_Heter_Lorentzian_DifPara***", results and XPPAUT codes see "Paper_MF_Izhikevich_Neuron/SA-WA_SameParaAsOnePop"

Fig.~\ref{fig:2pop-merge3} further illustrates the impact of the proportion of strongly adapting neurons in the network.  The boundary of the bursting region is shifted into the high mean intrinsic current region for a higher proportion [shown in (a)] and the rhythmic regime appears as $\kappa$ is increased [shown in (b)]. This indicates that stable bursting behavior is more likely in a network with a higher proportion of strongly adapting neurons. Fig.~\ref{fig:2pop-merge3} (c) and (d) shows a comparison between the full network dynamics and the reduced mean-field system. The reduced description captures the essential shape of the firing activity of the full network. A little discrepancy happens to the approximation of the mean adaption variable $\langle w \rangle$. This may be due to the failure of the assumption $\langle w|\eta \rangle \gg w_{\mathrm{jump}}$ during the derivation of dynamics of $\langle w \rangle$ (see \ref{app:dw}).

%%%%%%%%%%%%%%%%%%%%%%%%%%%%%%%%%%%%%

\section{DISCUSSION }\label{sec:discussion}

We have derived and validated a mean-field model for a network of heterogeneous Izhikevich neurons which display spike frequency adaptation through a recovery variable. 
We have further demonstrated that it is straightforward to apply our approach to multiple populations where the forces of adaptation, inhibition, and excitation interact. 
The mean-field models have exhibited qualitative and quantitative agreement with the full network.
Using bifurcation analysis, we have identified and characterized regimes of collective bursting that emerge given appropriate levels of adaptation, external stimulus and proportion of strongly adapting neurons.
The parameter values used in the numerical examples are  a nondimensionalization of parameter values fit to actual neuronal data collected in the literature.
Bifurcation analysis for the mean-field system can be used to make predictions about the biological networks being studied. For example, how to understand the emergence of bursting in the CA3b region of the hippocampus based on experimental findings of neurons which display different amounts of spike frequency adaptation \cite{Hemond2008}. The mean-field model we derived, (\ref{eq:2pop-mf-izh-p})-(\ref{eq:2pop-mf-s}), as an extension of that of \cite{Dur-e-Ahmad2012} and \cite{Nicola2013bif}, allows us to compute bifurcation manifolds and types for the network with different proportions of strongly and weakly adapting neurons and determine the impact of various parameters on the transition between the behavior of tonic firing and bursting in the actual neural network.

%\subsection{Validity analysis}
To assess the validity of our mean-field approximation, we examine all the assumptions that are imposed during the derivation. They are listed in order of appearance as follows.

1. $N \rightarrow \infty$, the thermodynamic limit.

In theory, the mean-field model is an exact description for the network of neurons in the thermodynamic limit. In the finite-size numerical experiments, the spread of the network variables around the mean narrows as the number of neurons increases, and thus gets closer to the dynamics of the mean-field model. One can see Fig.~\ref{fig:2pop-TimeEvolution} for a comparison, where $N_p = 8000$ for the strongly adapting neurons and $N_q = 2000$ for the weakly adapting neurons. The same trend can be seen in \cite[Fig.~6]{Ciszak2021}. 

%============
2. $\langle w|\eta \rangle \gg w_{\mathrm{jump}}$, the adaptation variable averaged over the network  
is much greater than the homogeneous after-spike jump value.

This assumption is required when deriving the differential equation of $\langle w\rangle$. The small discrepancy between the mean-field model and the population of strongly adapting neurons [see time series of $\langle w\rangle$ of population $p$ in Fig.~\ref{fig:2pop-TimeEvolution} (e) and (f), Fig.~\ref{fig:2pop-merge3} (c) and (d)] may result from a partial failure of the requirement. However, the mean-field description still captures the essential shape and frequency of the firing activity of the network.
Inclusion of high-order terms in the Taylor expansion in (\ref{eq:mean-w-Taylor-exp}) may improve the approximation. This will give rise to an extra term in the final mean-field equation for $\langle w \rangle'$.

%============
3. $\langle w|v,\eta \rangle = \langle w|\eta \rangle$, first-order moment closure assumption, also called the adiabatic approximation.

This assumption entails fast dynamics of the membrane potential. We could employ a high-order moment closure approximation, although we need to assess the value of the added effort in terms of the improvement of the accuracy of the resulting mean-field model \cite{Ly2007}. 
%especially since higher-order moment closure does not necessarily improve the accuracy  

%============
4. $\rho^v(t,v|\eta)=\frac{1}{\pi}\frac{x(t,\eta)}{\big [v-y(t,\eta)\big]^2+x^2(t,\eta)}$, the Lorentzian ansatz on the conditional density function. 

Derivation of the differential equations of the mean-field variables  $r(t)$ and $\langle v(t)\rangle$ started with writing out the population density function in the conditional form
$\rho (t,v,w,\eta) = \rho^w(t,w|v,\eta)\rho^v(t,v|\eta)\mathcal{L}(\eta)$
as shown in (\ref{eq:pop-density}). When we rewrite it as $\rho (t,v,w,\eta) = \rho^{vw}(t,v,w)\rho^\eta(t,\eta|v,w)$ and assume $\rho^{vw}(t,v,w)$ has the Lorentzian shape, we can use the method developed in the paper to obtain a different mean-field system.
\cite{Nicola2013hetero} shows how changing the expansion of the population density function can drastically change the resulting mean-field model. The expansion we used here corresponds to that used to develop "Mean-field system III" in \cite{Nicola2013hetero}.

%============
5. $ v_{\mathrm{peak}} = -v_{\mathrm{reset}}\rightarrow \infty $, limit of the resetting rule when neurons spike.

Parameter values used in the paper are based on actual neuronal data except the resetting values.
This choice facilitate analysis \cite{Izhikevich2007}, by linking the QIF model to the theta neuron model through the change of variable $v = \tan(\theta/2)$. %where the neuron is fired whenever $\theta$ crosses $\pi$.
Although not precisely biologically realistic, the theta model and its variants have been used in the literature to explore phenomena, such as rhythm generation \cite{KilpatrickErmentrout2011}, wave propagation in the cortex \cite{Byrne2020} and models for EEG \cite{byrne2022mean}. For the neuron model that exhibits a saddle-node on an invariant circle (SNIC) bifurcation, it is possible to reduce it to the theta model with adaptation, e.g. \cite{ermentrout1986parabolic,Ermentrout1996,KilpatrickErmentrout2011}. For the system not near a SNIC, but near some other bifurcation satisfying fairly general and biophysiologically plausible conditions, one can still 
%derive 
obtain the theta model with adaptation \cite{Izhikevich2000}.
When dealing with a biological network based on experimental data, we should treat the assumption carefully, as changing $v_\mathrm{peak}$ and $v_\mathrm{reset}$ can affect firing rates and estimation of the mean membrane potential.
In the numerical experiments, we choose $v_\mathrm{peak}
= -v_\mathrm{reset} = 200$ as in \cite{Dumont2017}, which is far away from the normal range of the membrane potential $v$.
We have found that different spiking thresholds lead to  some bias in terms of $\langle v(t)\rangle$ averaged over the full network, even if they meet the assumption requirement. 
Montbri\'o et al. \cite{Montbrio2015,Montbrio2020} attempted to address this
drawback by adding a refractory period to the network model, that is, the time for neurons taken from $v_\mathrm{peak}$ to infinity and minus infinity to $v_\mathrm{reset}$. This effectively makes the firing frequency and the mean membrane potential of the network match those of the theta model and hence the mean-field model.  
In addition, to avoid numerical delicacies near the spiking threshold, \cite{Pyragas2021} transformed the QIF form to the form of the theta neuron when performing the numerical simulation.

%============
6. $\mathcal{L}(\eta) =
\frac{1}{\pi}
\frac{\Delta_\eta}{(\eta - \bar \eta)^2 + \Delta_\eta^2}
$, distribution of the heterogeneous current.

Many parameters can be the sources of heterogeneity
in a network. Here we chose the intrinsic current $\eta$, but our approach could be applied to other
choices, such as 
%the external stimulus $I_\mathrm{ext}$,
the synaptic conductance $g_\mathrm{syn}$.  
In addition, the choice of the Lorentzian function is just a mathematical convenience. It can sharply reduce the complexity of the resulting mean-field model when evaluating the integrals in (\ref{eq:r-x}) and (\ref{eq:mean-v-y}). 
Unfortunately, the Lorentzian distribution is physically implausible since both its expected value and its variance are undefined.
Other distributions including Gaussian have been discussed in the literature. Particularly, if a distribution $\mathcal{L}(\eta)$ has $n$ complex-conjugate pole pairs, one can readily obtain the mean-field model consisting of $n$ complex-value ODEs in the form (\ref{eq:MF-complex}) \cite{Ott2008}.
Further, the papers \cite{Ott2009} and \cite{Montbrio2015} both pointed out that mean-field systems obtained using Lorentzian and Gaussian had qualitatively identical bifurcation structures. 
However, the paper \cite{Klinshov2021} claimed that the quantitative difference did matter in terms of both asymptotic and transient dynamics. It also showed how to compute the integrals in (\ref{eq:r-x}) and (\ref{eq:mean-v-y}) with the help of the rational approximation and residue theory  when $\mathcal{L}(\eta)$ is a Gaussian distribution.

For the numerical experiments, \cite{Montbrio2015} generated a
set of $N$ input currents that accurately reproduced a Lorentzian distribution and used the same set of input currents for all simulations with $N$ neurons.
Here we take a different approach. We generate the distribution by using the technique of inverse transform sampling. Specifically, for the $k$th neuron, we have
\begin{equation}
    \eta_k = 
    \bar \eta + 
    \Delta_\eta \tan \big (\pi (r_k-0.5) \big),
\end{equation}
where value of the cumulative distribution function $r$ is randomly 
sampled from the uniform distribution on $(0,1)$.
This is a basic method for pseudo-random number sampling from any probability distribution. The advantage is that the distribution of the heterogeneous currents is more realistic. The drawback is that the number of neurons $N$ involved in the simulation must be large enough to exhibit the Lorentzian shape in order to ensure the dynamics of  the network is consistent with that of the mean-field model. Additionally, numerical results obtained in every simulation are a little different since the current distributed to each neuron is different each time. As an example, \cite{Ciszak2021} showed similar accuracy of the mean-field results was observed by employing a network of $N = 10,000$ nodes with the deterministic generation rule and one of $N=1,000,000$ nodes with the random algorithm. 

%============
7. $\eta \in (-\infty, \infty)$, range of the heterogeneous current.

This assumption is adopted in evaluation of the integrals (\ref{eq:mean-v-y}) and (\ref{eq:r-x}) using the residue theorem.
For the neural network to be realistic in spite of this requirement, the distribution range of the heterogeneous parameter should be much wider than its half-width at half-maximum to achieve eq. (\ref{eq:rv-xy}).

%we need to choose a relatively wider distribution range of the heterogeneous parameter should be \textcolor{red}{much greater than $x(t,\eta)$ and $\Delta_\eta$}, the two half-widths at half-maximum of the distribution functions $\rho^v(t,v|\eta)$ and $\mathcal{L}(\eta)$ in (\ref{eq:LA}) and (\ref{eq:LA-para}), respectively.

In sum, all these seven requirements are not truly indispensable for applicability of the developed mean-field models. Some choices are a mere mathematical convenience, and insights gained 
from the macroscopic description are more generally applicable.

%==============================
%\subsection{Comparison with previous work}
To our knowledge, the mean-field models we derived  in this paper represent the first exact macroscopic description for a spiking neural network with adaptation, which does not rely on assumptions of explicit separation of time-scales, weak coupling or averaging. %or any other approximation. 
Nicola and Campbell \cite{Nicola2013bif,Nicola2013hetero} proposed a set of mean-field models for the homogeneous and heterogeneous Izhikevich network. A quasi-steady approximation was used by assuming the time scale of adaptation sufficiently large. The resulting mean-field model was a system of switching ordinary differential equations and the bifurcation theory of non-smooth systems had to be involved to perform further dynamical analysis. We used similar parameter values and thus our work can be directly compared.
The shape of the bursting region and dependence of bursting on various parameter is consistent, which is satisfying since the bursting mechanism in the underlying network model is the same.
Interestingly, we have found a novel mechanism for emergence of bursting in the neural network of two-coupled populations. Bursting was initiated by what appeared to be a saddle-node on an invariant
circle bifurcation or a homoclinic bifurcation [see Fig.~\ref{fig:2pop-merge1} (c)].
%Nicola et al. found that in a network with homogeneous input currents bursting was initiated by subcritical Hopf bifurcation, but in a heterogenoues it was initiated by a supercritical Hopf bifurcation. This is in contrast to our work where initiation occurs via subcritical Hopf \textcolor{red}{Fig. 4(c)?}.
%
Moreover, our results improve on those of \cite{Nicola2013bif,Nicola2013hetero} in the sense that the mean-field model gives a very accurate representation of the frequency of bursting. 
Gast et al. \cite{Gast2020NeuralComp} developed a smooth mean-field system for the QIF network with adaptation. The SFA mechanism acted additively to the dynamics of the membrane potential, just like the Izhikevich neuron. However, its adaptation variable was specifically expressed as a convolution of the membrane potential with an integral kernel. This treatment facilitates finding the closed set of mean-field equations, but lacks generality. Additionally, the adaptation dynamics in \cite{Gast2020NeuralComp} was also assumed to be slow enough that the variable was regarded as constant, finally leading to the same derivation process as that in \cite{Montbrio2015}. 
Recently, Bandyopadhyay et al. \cite{HH-mf2022} have derived a mean-field model for the network of the Hodgkin-Huxley neurons including the effect of ion-exchange between intracellular and extracellular environment. They use quasi-steady state assumptions and numerical fitting to approximate the Hodgkin-Huxley model by a piecewise-defined QIF model and then apply the Montbri\'o approach. However, the incorporation of the ion-exchange dynamics is ad hoc. 
By comparison, we provide explicit and solid mathematical foundations underlying the derivation process, that show how to incorporate such additional variables into a mean-field model.
%that would contribute to deeply understanding the accuracy and efficacy of our proposed mean-field model, and deliver accessible tools to investigate brain functions and dysfunctions.

%============================
Interaction between fast and slow processes in a network of spiking neurons can induce much richer  dynamics, especially the emergence of population bursting activity and the resulting spike synchronization.
Those regimes are of interest to describe both normal and pathological neural network dynamics. 
The mean-field models developed in this paper provide a tractable and reliable tool to investigate the underlying mechanism of brain function from the perspective of theoretical neuroscience. For example, Rich et al. \cite{Rich2020} performed a computational analysis for a network of $500$ Izhikevich neurons to explore a novel hypothesis about the seizure initiation.  We expect that theoretical analysis through our mean-field models can provide reasonable interpretation for numerical results in \cite{Rich2020}. 
In addition, the impact of time delay, gap-junctions, realistic network topology, may be considered within the same framework allowing for the application to more biologically realistic network models. %and further bifurcation analysis are expected.

%%%%%%%%%%%%%%%%%%%%%%%%%%%%%%%%%%%%%

\section*{ACKNOWLEDGMENTS}
This work benefited from the support of the Natural Science and Engineering Research Council of Canada.

%%%%%%%%%%%%%%%%%%%%%%%%%%%%%%%%%%%%%

\appendix
\section{Derivation of dynamics of the mean adaptation variable} \label{app:dw}

The idea in this section is from \cite{Nicola2013hetero}. The continuity equation (\ref{eq:CE-general}) and (\ref{eq:flux-general}) yields
\begin{align}\label{eq:mean-w-1}
\langle w \rangle' 
&= 
\int_\eta \int_v \int_w
w 
\frac{\partial}{\partial t}\rho(t,v,w,\eta)
\mathrm{d}w
\mathrm{d}v
\mathrm{d}\eta \nonumber\\
&=
-\int_\eta \int_v \int_w
w \left (
\frac{\partial \mathcal{J}^v}{\partial v}
+
\frac{\partial \mathcal{J}^w}{\partial w}
\right) 
\mathrm{d}w
\mathrm{d}v
\mathrm{d}\eta 
\end{align}
Next, we evaluate the two terms on the right hand side of (\ref{eq:mean-w-1}), respectively. For the first term, we apply integration by parts and change the order of integration as needed, then obtain
\begin{align}
\mathrm{Term} 1
&=
\int_\eta \int_v \int_w
w 
\frac{\partial \mathcal{J}^v}{\partial v}
\mathrm{d}w
\mathrm{d}v
\mathrm{d}\eta \nonumber \\
&=
\int_\eta \int_w \int_v
w
\mathrm{d}\mathcal{J}^v(t,v,w,s,\eta)
\mathrm{d}w
\mathrm{d}\eta \nonumber \\
&=
\int_\eta \int_w
w
\mathcal{J}^v(t,v,w,s,\eta)|_{\partial v}
\mathrm{d}w
\mathrm{d}\eta \nonumber \\
&=
\int_\eta \int_w
w \Big (
\mathcal{J}^v(t,v_{\mathrm{peak}},w,s,\eta)
-
\mathcal{J}^v(t,v_{\mathrm{reset}},w,s,\eta)
\Big )
\mathrm{d}w
\mathrm{d}\eta \nonumber \\
&=
\int_\eta \int_w
w \Big (
\mathcal{J}^v(t,v_{\mathrm{peak}},w,s,\eta)
-
\mathcal{J}^v(t,v_{\mathrm{peak}},w-w_{\mathrm{jump}},s,\eta)
\Big )
\mathrm{d}w
\mathrm{d}\eta 
\end{align}
Assuming $\langle w|\eta \rangle \gg w_{\mathrm{jump}}$, we apply a Taylor expansion and integration by parts and have
\begin{align}\label{eq:mean-w-Taylor-exp}
\mathrm{Term} 1
&=
\int_\eta  \int_w
w
\left (
w_{\mathrm{jump}}
\frac{\partial}{\partial w}\mathcal{J}^v(t,v_{\mathrm{peak}},w,s,\eta)
+ 
\mathcal{O}(w^2_{\mathrm{jump}})
\right )
\mathrm{d}w
\mathrm{d}\eta \nonumber \\
&=
\int_\eta w_{\mathrm{jump}} 
\mathrm{d}\eta 
\int_w
w \mathrm{d}\mathcal{J}^v(t,v_{\mathrm{peak}},w,s,\eta)
\mathrm{d}w 
+
\mathcal{O}(w^2_{\mathrm{jump}})
\nonumber \\
& \approx
\int_\eta w_{\mathrm{jump}}
\mathrm{d}\eta
\left (
w \mathcal{J}^v(t,v_{\mathrm{peak}},w,s,\eta)|_{\partial w}
-
\int_w
\mathcal{J}^v(t,v_{\mathrm{peak}},w,s,\eta)
\mathrm{d}w
\right )
\end{align}
Then, we assume the flux to be vanishing on the boundary $\partial w$, yielding 
\begin{equation}
\mathrm{Term} 1
= 
- \int_\eta \int_w
w_{\mathrm{jump}}
\mathcal{J}^v(t,v_{\mathrm{peak}},w,s,\eta)  
\mathrm{d}w
\mathrm{d}\eta.
\end{equation}
%
%=====================
Similarly, for the second term of (\ref{eq:mean-w-1}), we have
\begin{align}
\mathrm{Term} 2
&=
\int_\eta \int_v \int_w
w 
\frac{\partial \mathcal{J}^w}{\partial w}
\mathrm{d}w
\mathrm{d}v
\mathrm{d}\eta \nonumber \\
&=
\int_\eta \int_v \int_w
w
\mathrm{d}\mathcal{J}^w 
\mathrm{d}v
\mathrm{d}\eta \nonumber \\
&= 
\int_\eta \int_v
\left (
w\mathcal{J}^w|_{\partial w}
-
\int_w  \mathcal{J}^w \mathrm{d}w
\right )
\mathrm{d}v
\mathrm{d}\eta \nonumber \\
&=
- \int_\eta \int_v \int_w
\mathcal{J}^w
\mathrm{d}w
\mathrm{d}v
\mathrm{d}\eta \nonumber \\
&=
-\int_\eta \int_v \int_w
G^w(t,v,w)\rho(t,v,w,\eta)
\mathrm{d}w
\mathrm{d}v
\mathrm{d}\eta \nonumber \\
&=
-\langle G^w(t,v,w) \rangle.
\end{align}
Finally, we obtain the differential equation of the mean adaptation variable in terms of flux given by
\begin{equation}
\langle w \rangle' 
=
\langle G^w(t,v,w) \rangle
+
\int_\eta \int_w
w_{\mathrm{jump}}
\mathcal{J}^v(t,v_{\mathrm{peak}},w,s,\eta)  
\mathrm{d}w
\mathrm{d}\eta.
\end{equation}

%%%%%%%%%%%%%%%%%%%%%%%%%%%%%%%%%%%%%

\section{Dimensional Izhikevich network and its nondimensionalization}\label{app:nondim}

The dimensional form of the network of Izhikevich neurons is the same as that described by \cite{Nicola2013bif}, and is given by
\begin{subequations}\label{eq:network-Izh-dim}
\begin{align}
    C\frac{d V_k}{dT} 
    &=
    k_1(V_k-V_T)(V_k-V_R) - W_k + I_{\mathrm{app},k} + G_{\mathrm{syn}}s(E_r - V_k) \\
    \tau_W \frac{d W_k }{dT} 
    &=
    \beta(V_k-V_R) - W_k \\
    \frac{ds}{dT}
    &=
    -\frac{s}{\tau_{\mathrm{syn}}} 
    +
     \frac{S_{\mathrm{jump}}}{N}
    \sum_{k=1}^{N}
    \sum_{t_{k,j}<t}
    \delta(t-t_{k,j}) \\
    \mathrm{if} \;
    &
    V_k \geq V_{\mathrm{peak}}, 
    \mathrm{then} \;
    V_k \leftarrow V_{\mathrm{reset}}\;
    \mathrm{and}\;
    W_k \leftarrow W_k + W_{\mathrm{jump}},
\end{align}
\end{subequations}
where $k=1, 2, \dots N$. Parameters are interpreted in Table~\ref{tab:para-dimensional} and values are chosen to fit hippocampal CA3 pyramidal neuron data.

\begin{table}[ht!]
\centering\footnotesize
\begin{threeparttable}
\caption{Parameters for the dimensional Izhikevich network based on \cite{Nicola2013bif,Nicola2013hetero}}
\label{tab:para-dimensional}
\begin{tabular}{l r l}
  \toprule
  Parameter & Value & Description\\
  \midrule
  C 
  & $250\ \mathrm{pF}$ 
  & Membrane capacitance\\
  $k_1$ 
  & $2.5\ \mathrm{nS/mV}$
  & Scaling factor\\
  $V_R$
  & $-65\ \mathrm{mV}$
  & Resting membrane potential\\
  $V_T$
  & $V_R + 40 - \frac{\beta}{k_1} = -24.6 \ \mathrm{mV}$
  & Threshold potential\\
  $G_{\mathrm{syn}}$
  &
  $200 \ \mathrm{nS}$ 
  & Synaptic conductance\\
  $E_r$
  &
  $0 \ \mathrm{mV}$
  & Reversal potential\\
  $\beta$
  &
  $-1 \ \mathrm{nS}$
  &  Scaling factor\\
  $\tau_W$
  &
  $200 \ \mathrm{mS}$ 
  & Adaptation time constant\\
  $\tau_{\mathrm{syn}}$
  &
  $4 \ \mathrm{mS}$ 
  & Time constant of the gating variable\\
  $S_{\mathrm{jump}}$
  &
  $0.8$ 
  & Coupling strength\\
  $W_{\mathrm{jump}}$
  &
  $200 \ \mathrm{pA}$ 
  & After-spike jump size of the variable $W$\\
\bottomrule
\end{tabular}
\end{threeparttable}
\end{table}  

\begin{figure}[ht!]
\centering
\includegraphics[width=0.9\textwidth]{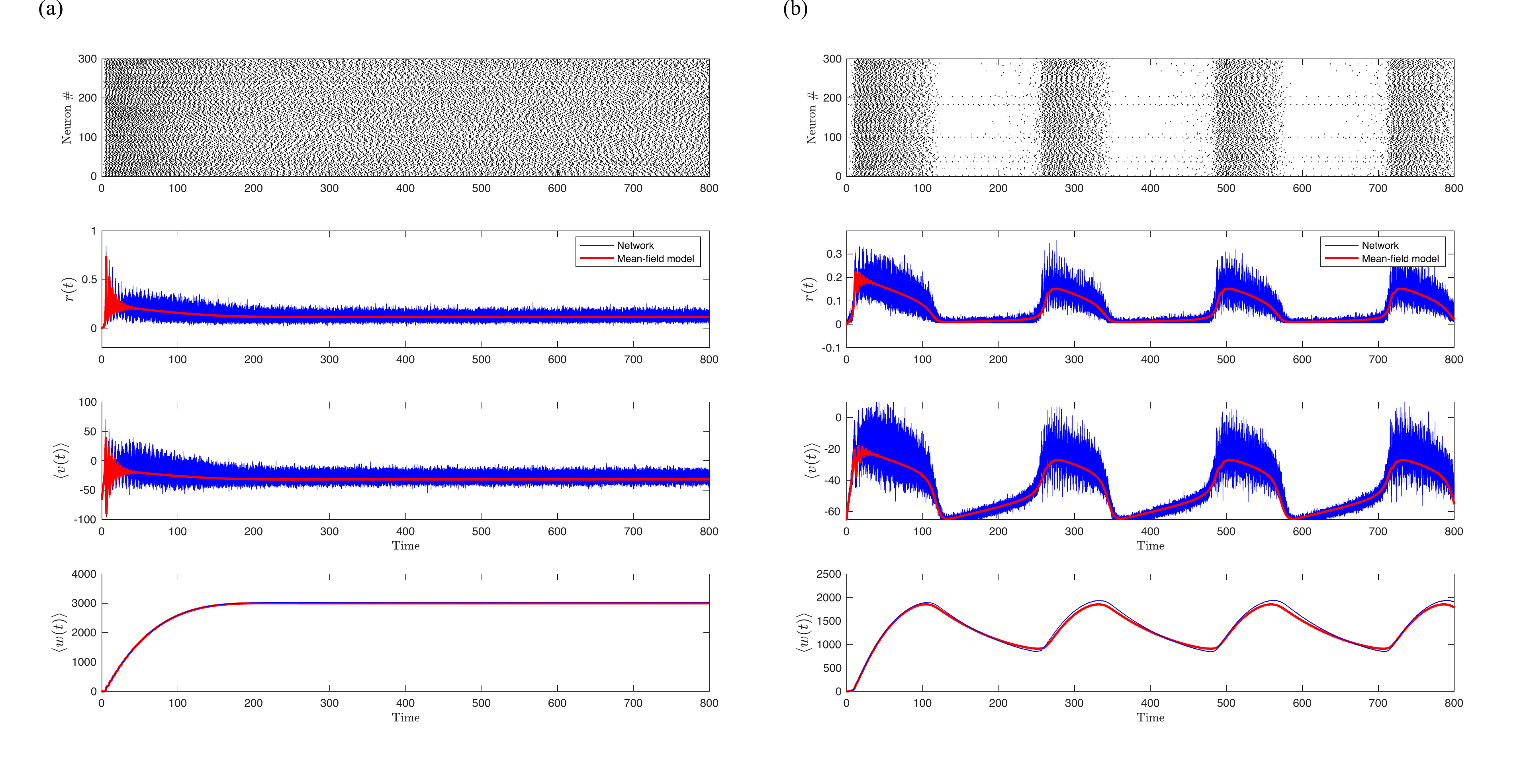}
\caption{Time evolution for the dimensional network of Izhikevich neurons (\ref{eq:network-Izh-dim}) (shown in blue) and the corresponding mean-field model (shown in red) when the mean intrinsic current is $2641 \ \mathrm{pA}$ and $1268 \ \mathrm{pA}$, respectively. The two plots correspond to Fig.~\ref{fig:1pop-merge1} (c) and (d) for the dimensionless systems.}
\label{fig:1pop-dim}
\end{figure}

\begin{figure}[ht!]
\centering
\includegraphics[width=0.9\textwidth]{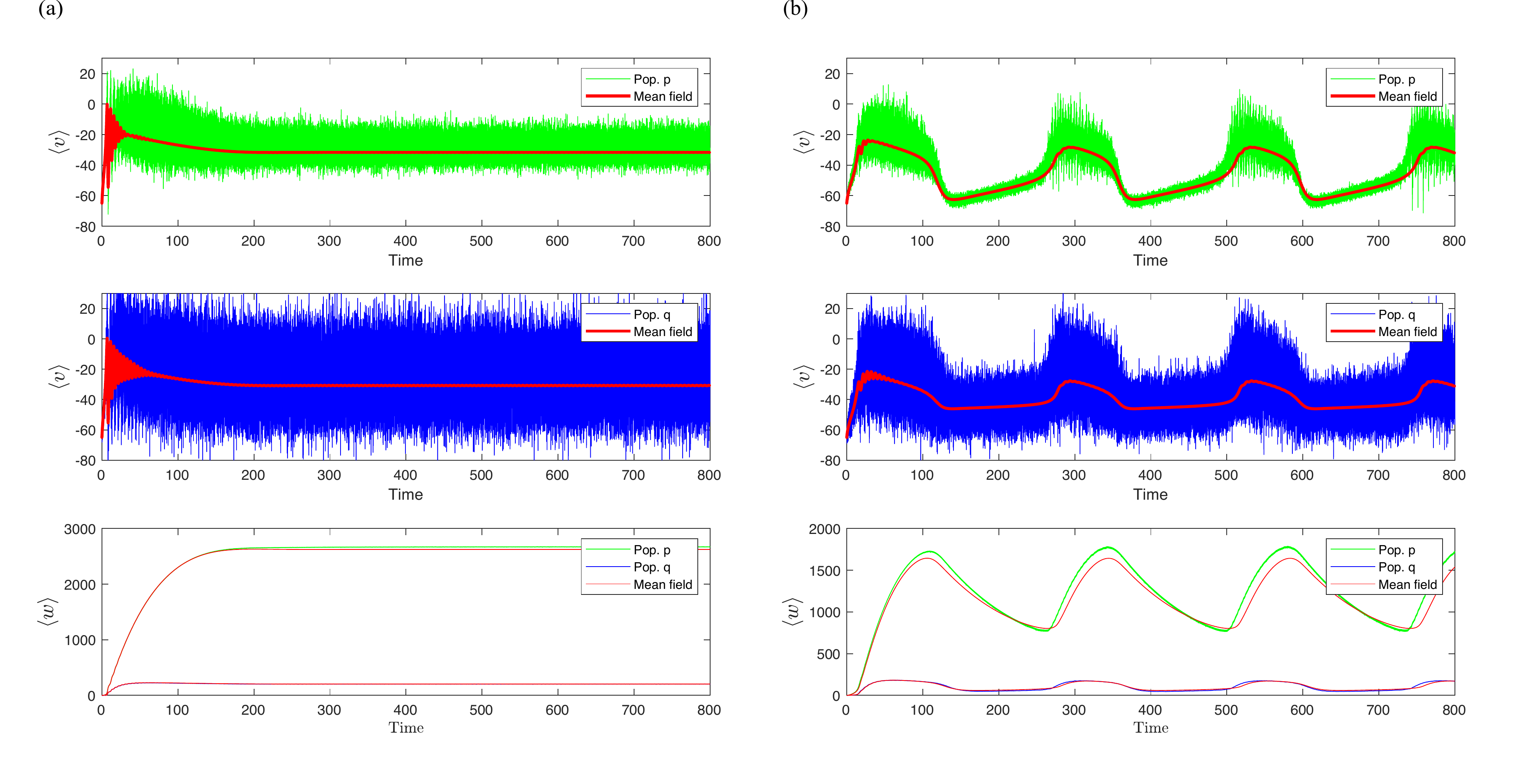}
\caption{Time evolution for the dimensional network of two-coupled populations of Izhikevich neurons and the corresponding mean-field model when both mean intrinsic currents are $1091 \ \mathrm{pA}$ and $845 \ \mathrm{pA}$, respectively. The two plots correspond to Fig.~\ref{fig:2pop-TimeEvolution} for the dimensionless systems.}
\label{fig:2pop-vw-dim}
\end{figure}

The corresponding dimensionless form of the network model is
\begin{align}
   \frac{d v_k}{dt}  
    &=
    v_k(v_k - \alpha)
    -
    w_k
    +
    I_k 
    + g_{\mathrm{syn}}s(e_r - v_k)
    \nonumber \\
    \frac{d w_k}{dt} 
    &=
    a(b v_k - w_k)
    \nonumber \\
    \frac{ds}{dt} 
    &=
    -\frac{s}{\tau_s}
    +
    \frac{s_{\mathrm{jump}}}{N}
    \sum_{k=1}^{N}
    \sum_{t_{k,j}<t}
    \delta(t-t_{k,j})
\end{align}
where $I_k$ corresponds to $\eta_k + I_{\mathrm{ext}}$ in (\ref{eq:network-Izh}a).
Critical transformations utilized in the process of nondimensionalization are listed as follows,
\begin{equation*}
    V_k -V_R 
    = V_k + |V_R| 
    = |V_R|
    \left(
    1 + \frac{V_k}{|V_R|}
    \right )
    = |V_R|v_k, \quad v_k = 1 + \frac{V_k}{|V_R|}
\end{equation*}

\begin{equation*}
    C \frac{dV_k}{d T}
    =
    \frac{C}{k_1|V_R|} \frac{d v_k}{d T}
    = 
    \frac{C}{k_1|V_R|} \frac{d t}{d T} \frac{dv_k}{d t}
    = 
    \frac{dv_k}{d t}, \quad  t = \frac{k_1 |V_R|}{C} T
\end{equation*}

\begin{equation*}
    \frac{ds}{d T} = \frac{d t}{d T}\frac{ds}{d t}
    = \frac{k_1 |V_R|}{C} \frac{ds}{d t}
\end{equation*}
The scaling relationship with the dimensional system is shown in \ref{tab:scaling}. 
%
%To help understand the biological neural network,
As should be expected, numerical simulation of the dimensional network models and the corresponding mean-field approximations gives the same results as the corresponding dimensionless systems. This  is shown in  Fig.~\ref{fig:1pop-dim} and \ref{fig:2pop-vw-dim}, which  correspond to Fig.~\ref{fig:1pop-merge1} for the dimensionless network of single population and Fig.~\ref{fig:2pop-TimeEvolution} for the one with two populations, respectively.

\begin{table}[ht!]
\centering\footnotesize
\begin{threeparttable}
\caption{Scaling relations between the dimensionless and dimensional neural network}
\label{tab:scaling}
\renewcommand*{\arraystretch}{1.5}
\begin{tabular}{l l l l}
\toprule
$v_k = 1+\frac{V_k}{|V_R|} $
& $w_k = \frac{W_k}{k_1|V_R|^2}$ 
& $s   = s$
& $T = \frac{C}{k_1 |V_R|} t$ 
\\ 
$v_{\mathrm{peak}} =
    1 + \frac{V_{\mathrm{peak}}}{|V_R|}$
& $v_{\mathrm{reset}} = 
    1 + \frac{V_{\mathrm{reset}}}{|V_R|}$ 
& $\alpha =
    1 + \frac{V_T}{|V_R|}$ 
& $g_{\mathrm{syn}} =
\frac{G_{\mathrm{syn}}}{k_1|V_R|}$
\\
$a =
\left(
\frac{\tau_W k_1|V_R|}{C}
\right)^{-1} $
& $b =\frac{\beta}{k_1|V_R|} $ 
& $s_{\mathrm{jump}} =
S_{\mathrm{jump}}\frac{C}{k_1|V_R|}$
& $w_{\mathrm{jump}}= 
\frac{W_{\mathrm{jump}}}{k_1 |V_R|^2}$
\\
$e_r = 1 + \frac{E_r}{|V_R|} $
& $I =
\frac{I_{\mathrm{app}}}{k_1|V_R|^2} $
& $\tau_s = \frac{\tau_{\mathrm{syn}}k_1|V_R|}{C}$ \\
\bottomrule
\end{tabular}
\end{threeparttable}
\end{table}

% matlab codes from "1_IzhikevichNeuron_Heter_Lorentzian_burst***", "results in the paper"

% matlab codes from "2_Izhikevich_StrongWeakAdapt_Heter_Lorentzian_DifPara***", "EP_mu=0.18/fig_plot3_vw_dim", "PO_mu=0.08"

%%%%%%%%%%%%%%%%%%%%%%%%%%%%%%%%%%%%%
%\section*{References}

\bibliography{mybib}

\end{document}